\documentclass[letterpaper,twocolumn,fleqn]{article}

\usepackage{ist}
\usepackage{graphicx}
\usepackage{pifont}
\usepackage{algorithm}
\usepackage{mathtools}
\usepackage{algpseudocode}
\usepackage{pbox}
\usepackage{footmisc}
\usepackage{subcaption}
\usepackage{xr}
\usepackage{makecell}
\usepackage{tikz}
\usepackage{comment}
\usepackage{color,colortbl}
\usepackage{makecell}
\usepackage{booktabs}
\usepackage{float}
\usepackage{array}
\usepackage{cite}
\usepackage{balance}

\DeclareMathAlphabet{\altmathcal}{OMS}{cmsy}{m}{n}

\usepackage{setspace}
\usepackage{graphicx}
\usepackage{multirow,multicol}
\usepackage[]{amsmath}
\usepackage{amsbsy}
\usepackage{amssymb}
\usepackage{amsfonts}
\usepackage{pifont}
\usepackage{balance}
\usepackage{fancyvrb}
\usepackage[framemethod=TikZ]{mdframed}

{\begin{list}               
    {$\bullet$ \hfill}{
        \setlength{\leftmargin}{\parindent}
        \setlength{\parsep}{0.04\baselineskip}
        \setlength{\itemsep}{0.5\parsep}
        \setlength{\labelwidth}{\leftmargin}
        \setlength{\labelsep}{0em}}
    }
{\end{list}}

\newcommand{\boxedeg}[1]{
	\begin{mdframed}[roundcorner=5pt,middlelinewidth=2pt,backgroundcolor=yellow!10]
		\vspace{-1ex}
		#1
	\end{mdframed}
}

\providecommand{\eref}[1]{\eqref{#1}}  
\providecommand{\cref}[1]{Chapter~\ref{#1}}

\providecommand{\fref}[1]{Figure~\ref{#1}}

\providecommand{\R}{\ensuremath{\mathbb{R}}}

\providecommand{\E}{\ensuremath{\mathbb{E}}}

\providecommand{\bydef}{\overset{\text{def}}{=}}

\renewcommand{\vec}[1]{\ensuremath{\boldsymbol{#1}}}

\providecommand{\calA}{\altmathcal{A}}

\providecommand{\calE}{\altmathcal{E}}

\providecommand{\calG}{\altmathcal{G}}
\providecommand{\calH}{\altmathcal{H}}

\providecommand{\calR}{\altmathcal{R}}

\providecommand{\mA}{\mathbf{A}}

\providecommand{\mH}{\mathbf{H}}

\providecommand{\mM}{\mathbf{M}}

\providecommand{\mZ}{\mathbf{Z}}

\providecommand{\vb}{\mathbf{b}} 

\providecommand{\vd}{\mathbf{d}}

\providecommand{\vh}{\mathbf{h}}

\providecommand{\vn}{\mathbf{n}}

\providecommand{\vt}{\mathbf{t}}
\providecommand{\vu}{\mathbf{u}}

\providecommand{\vx}{\mathbf{x}}
\providecommand{\vy}{\mathbf{y}}

\providecommand{\valpha}{\vec{\alpha}}
\providecommand{\vbeta}{\vec{\beta}}

\providecommand{\vtheta}{\vec{\theta}}

\providecommand{\vphi}{\vec{\phi}}
\providecommand{\vvarphi}{\vec{\varphi}}

\providecommand{\vpsi}{\vec{\psi}}

\providecommand{\vxhat}{\boldsymbol{\widehat{x}}}

\newcommand{\argmin}[1]{\mathop{\underset{#1}{\mbox{argmin}}}}

\definecolor{Gray}{gray}{0.9}
\newcolumntype{g}{>{\columncolor{Gray}}c}

\title{Computational Image Formation: Simulators in the Deep Learning Era}

\author{Stanley~H.~Chan; Purdue University, West Lafayette, United States}

\date{} 

\begin{document}

\maketitle

\thispagestyle{empty} 


\begin{abstract}
At the pinnacle of computational imaging is the co-optimization of camera and algorithm. This, however, is not the only form of computational imaging. In problems such as imaging through adverse weather, the bigger challenge is how to accurately simulate the forward degradation process so that we can synthesize data to train reconstruction models and/or integrating the forward model as part of the reconstruction algorithm. This article introduces the concept of computational image formation (CIF). Compared to the standard inverse problems where the goal is to recover the latent image $\vx$ from the observation $\vy = \calG(\vx)$, CIF shifts the focus to designing an approximate mapping $\calH_{\vtheta}$ such that $\calH_{\vtheta} \approx \calG$ while giving a good image reconstruction result. The word ``computational'' highlights the fact that the image formation is now replaced by a numerical simulator. While matching the mother nature remains an important goal, CIF pays even greater attention on strategically choosing an $\calH_{\vtheta}$ so that the reconstruction performance is maximized.

The goal of this article is to conceptualize the idea of CIF by elaborating on its meaning and implications. The first part of the article is a discussion on the four attributes of a CIF simulator: accurate enough to mimic $\calG$, fast enough to be integrated as part of the reconstruction, provides a well-posed inverse problem when plugged into the reconstruction, and differentiable to allow backpropagation. The second part of the article is a detailed case study based on imaging through atmospheric turbulence. A plethora of simulators, old and new ones, are discussed. The third part of the article is a collection of other examples that fall into the category of CIF, including imaging through bad weather, dynamic vision sensors, and differentiable optics. Finally, thoughts about the future direction and recommendations to the community are shared.
\end{abstract}

\section{Introduction}
\begin{figure}[t]
	\centering
	\includegraphics[width=\linewidth]{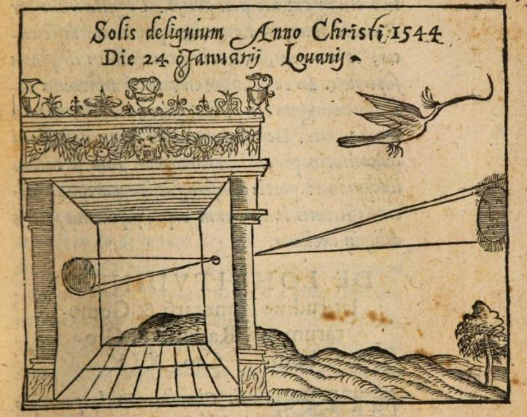}
	\caption{A schematic drawing of the pinhole camera made by Gemma Frisius in 1544 \cite{Frisius_1558}.}
	\label{fig: pinhole}
\end{figure}

The cameras we use today are largely a variant of the pinhole camera which, according to some scientists, can be traced back to nomadic tribes of North Africa thousands of years ago. A pinhole camera is easy to understand. Many of you have probably seen Gemma Frisius's drawing \cite{Frisius_1558} (See \fref{fig: pinhole}): Light is emitted from the source, propagating through the medium, and finally arriving at a tiny pinhole. Assuming that light travels along a straight line, a scaled and inverted image is formed on the screen. Today's cameras are certainly more complicated but they arguably follow the same principle, just with slightly more advanced optical elements to bend the light and a sensor to record the image intensity.

Over the past century, imaging has evolved to a much more diverse form now which we call computational imaging \cite{Bouman_2022_Book, Bhandari_2022_Book}. Unlike pinhole cameras that aim to produce a sharp image, a computational imaging system co-designs the sensing unit with the algorithm. The signal generated at the junction of sensor and algorithm may not necessarily be a sharp and clean image. It can be an unrecognizable signal, but with the magic of the algorithm, we can recover the image.

But computational imaging should not be bounded to the co-design of a \emph{hardware} and algorithm. In this paper, I want to introduce the concept of \emph{computational image formation} (CIF). CIF, in my mind, is a branch in computational imaging with the focus on selecting and optimizing the image formation process. This is perhaps a confusing (or even redundant) idea, because the image formation process is defined by either mother nature or hardware. If it is nature, e.g., shot noise due to the Poisson arrival process, we have a very precise equation describing how images are formed. If it is hardware, we have already seen many great examples such as coded aperture, coded exposure, CT, MRI, etc. So, what does computational image formation mean?

The core substance of CIF is a \emph{simulator}, a differentiable, fast, and accurate simulator that can be integrated into the image \emph{reconstruction} framework. You may wonder: ``A simulator? We have a ton of simulators in physics, and they are great. Why do we need \emph{your} simulator?'' The attributes of the simulator(s) I like to discuss here are quite different from what a physics simulator looks like. In physics, a simulator's only job is to mimic nature or processes we can observe but cannot control. In CIF, the focus of the simulator is not about matching mother nature unconditionally, but about maximizing the image quality of the image reconstruction algorithm. To this end, a simulator in CIF needs to have four properties:
\begin{enumerate}
\setlength\itemsep{0ex}
	\item \textbf{Accurate}. It should be accurate enough to mimic the true image formation process. The level of precision is determined by the image reconstruction goal.
	\item \textbf{Fast}. It needs to be fast enough so that it can be used to synthesize training data and/or integrated in the image reconstruction loop.
	\item \textbf{Benefits reconstruction}. It needs to improve the performance of the image reconstruction. To do so, the simulator's parameters can be updated during inference.
	\item \textbf{Differentiable}. It needs to be differentiable so that the reconstruction neural network can be trained while having the simulator in the loop.
\end{enumerate}

My goal in this paper is to conceptualize CIF and invite discussions from colleagues. I want to explain why I think CIF is important in our time. I will use atmospheric turbulence as a case study, but a few other examples will also be introduced to support my discussions. I will conclude the paper with some thoughts about moving the field forward.

\section{Concept of Computational Image Formation}

\subsection{Pinhole Camera}
In a pinhole camera\footnote{I am making a distinction between a conventional camera and a computational camera. But since the spellings of the two appear similar, I call the former as a pinhole camera. }, we can think of the camera being a passive device. If we use $\vx \in \R^d$ to denote the ground truth image in the object plane, and $\vy \in \R^m$ to denote the observed image in the image plane, the camera can be mathematically described as a mapping $\calH$ from $\vx$ to $\vy$:
\begin{equation}
	\underset{\text{observed image}}{\underbrace{\vy}} = \underset{\text{camera}}{\underbrace{\calH}} \;\;\; (\underset{\text{true image}}{\underbrace{\vx}}).
\end{equation}
The camera model $\calH$ can involve lenses, color filter arrays, image sensors, etc.

A post-processing image reconstruction algorithm for a pinhole camera is to find the best estimate $\widehat{\vx}$ by solving a certain optimization (e.g., maximum likelihood estimation or maximum-a-posteriori estimation) or running it through a neural network. For generality, I simply call the reconstruction as an operator $\calR$:
\begin{equation}
	\widehat{\vx} = \underset{\text{reconstruction}}{\underbrace{\calR(\vy, \calH)}}.
\end{equation}
The reconstruction method requires two inputs: the measurement $\vy$ and the camera model $\calH$. As a quick example, for those who do image deconvolution, $\calR$ can be a total variation minimization:
\begin{equation}
\widehat{x} = \calR(\vy,\calH) = \argmin{\vx} \; \|\calH(\vx)-\vy\|^2 + \lambda \|\vx\|_{
\text{TV}},
\end{equation}
where $\calH(\vx)$ denotes the blurring operation, and $\|\cdot\|_{\text{TV }}$ denotes the total variation norm. We can come up with many other examples which I shall skip for brevity.

The key observation of the above equations is that in a pinhole camera, we perform a capture-then-reconstruct operation. The camera is \emph{pre-defined}. While we use the model $\calH$ during reconstruction, we never send any \emph{feedback} signal to $\calH$. Furthermore, the design of $\calH$ is separated from the reconstruction. If we ask a camera engineer to build a camera $\calH$, he/she will never ask if we are using total variation minimization or a generative adversarial network for image reconstruction. With probability one, his/her goal is to make a $\calH$ that produces the ideal image, for good reasons.

\subsection{Computational Camera}
In a computational camera, as many of you know, the above isolation of the camera and the algorithm is replaced by a co-design philosophy. To accomplish this goal, I \emph{parameterize} $\calH$ with a \emph{state vector} $\vtheta$ so that the model becomes $\calH_{\vtheta}$.

\begin{figure}[t]
	\centering
	\includegraphics[width=\linewidth]{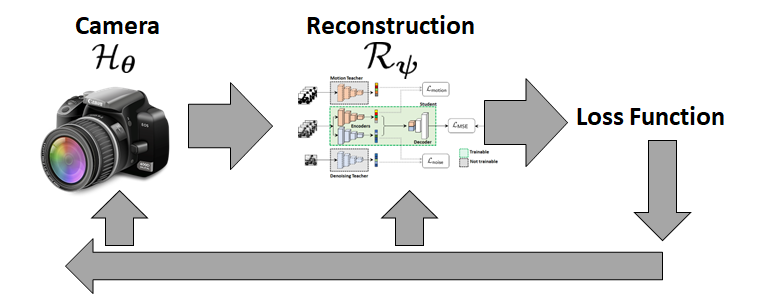}
	\caption{In a computational camera, the goal is to configure the camera parameters and the reconstruction's parameters so that the quality of the final reconstructed image is maximized.}
	\label{fig: Ch0 CompImagA}
\end{figure}

The presence of the state vector $\vtheta$ makes the overall design interesting, as shown in \fref{fig: Ch0 CompImagA}. Instead of solving a two-stage feed forward problem (design $\calH$ and then design $\calR$), the reconstructed image $\vxhat$ will be used to guide the design of $\calH$ so that we can close the loop. By parameterizing $\calR$ as $\calR_{\vpsi}$, the end-to-end design can be written as
\begin{align}
	(\widehat{\vpsi}, \widehat{\vtheta})
	&= \argmin{\vpsi,\vtheta} \;\; \E_{\vx} \Big[ \text{ReconLoss}\Big( \underset{\widehat{\vx}}{\underbrace{\calR_{\vpsi}( \vy, \; \calH_{\vtheta} )}}, \vx \Big) \Big],
	\label{eq: comp camera 1}
\end{align}
where $\text{ReconLoss}(\widehat{\vx},\vx)$ denotes the reconstruction loss between the predicted image $\widehat{\vx}$ and the ground truth $\vx$. I keep the definition of the loss function vague because the choice of the loss function depends on applications. It can be the squared loss, the cross-entropy loss, or any other loss that would make sense for the application.

\boxedeg{
	\vspace{2ex}
	
\textbf{Example 1}. (Coded aperture / lensless imaging) Consider a coded aperture camera where we are interested in recovering the full signal $\vx \in \R^d$ from a coded measurement $\vy \in \R^m$. The coding scheme we use is a fat matrix $\mH \in \R^{m \times d}$ where $m \ll d$ so that the measured signal
\begin{equation*}
	\vy  = \calH_{\vtheta}(\vx)\bydef  \underset{\text{encoded signal}}{\underbrace{\mH \vx}} \quad + \underset{\text{noise term}}{\underbrace{\vn}}
\end{equation*}
has a lower dimension than the true signal $\vx$.

\begin{center}
\includegraphics[width=\linewidth]{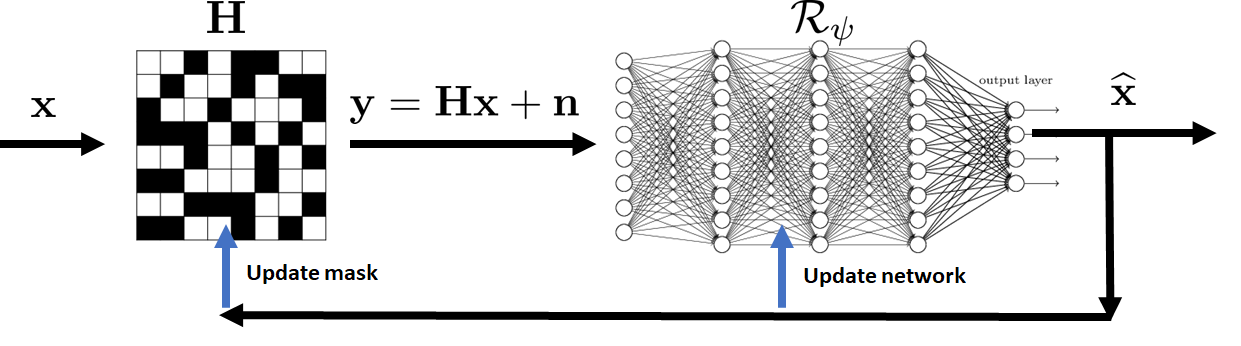}
\end{center}

To reconstruct the signal we create a reconstruction algorithm (e.g., a neural network) that does
\begin{equation*}
	\vxhat = \calR_{\vpsi}( \vy ).
\end{equation*}
Here, we can think of $\vpsi$ being the weights of the neural network.

To train the neural network and simultaneously find the optimal sensing matrix $\mH$, we perform the joint optimization
\begin{equation*}
	\widehat{\mH}, \widehat{\vpsi} = \argmin{\mH, \vpsi} \;\; \E_{\vx, \vn} \Big[ \text{ReconLoss}( \calR_{\vpsi}( \mH\vx + \vn), \vx ) \Big].
\end{equation*}
The expectation is taken with respect to both the signal $\vx$ and noise $\vn$ because they are random.

If we consider the hardware feasibility, we can further pose constraints on $\mH$. For example, we can require $\mH$ to be binary so that it can be implemented through the digital micro-mirror devices (DMD).
}

\subsection{Nature and Simulator}
The biggest difference between a computational camera and computational image formation (CIF) is the presence of a \emph{simulator}. To explain the idea it would be useful to consider the problem of imaging through an undesirable environment, e.g., fog, haze, turbulence, or scattering medium.

\textbf{Nature}. The subject of CIF is often concerned with \emph{nature}. For the time being let me just call it a general degradation process, denoted by $\calG$:
\begin{equation}
	\underset{\text{observed image}}{\underbrace{\vy}} = \underset{\text{nature}}{\underbrace{\calG}} \;\;\; (\underset{\text{true image}}{\underbrace{\vx}}).
\end{equation}
The degradation process is unknown to us. We have no idea of how each point in the object plane is mapped to a digital value in the image plane. We can in theory run procedures such as ray tracing to find out how each ray is distorted. But there will be infinitely many rays to trace and so realistically we can never perform such calculations. In some situations such as random scattering medium, we cannot even trace rays because it is impossible to know how each molecule affects the light.

\boxedeg{
\vspace{2ex}

\textbf{Example 2}. (Modeling haze) One of the landmark papers in imaging through weather is the dark channel prior by He et al. published in CVPR 2009 \cite{He_2009_darkchannelprior}, although the modeling can be traced back to Fattal \cite{Fattal_2008}, Rossum and Nieuwenhuizen \cite{Rossum_1999_scattering} and Koschmieder \cite{Koschmieder_1924}.

When light propagates through a scattering medium, the water molecules along the propagation path will cause attenuation of the light. The exact image formation process is governed by nature, and is unknown. We denote it as $\calG$.

To simulate the image formation, people use the so-called radiative transport equation by modeling the overall degradation as a sum of the airlight term and the direct attenuation term. This gives us a simulator $\calH_{\vtheta}$:
\begin{equation*}
\calH_{\vtheta}(\vx) = \vx \odot \vt + \mA (\mathbf{1}-\vt),
\end{equation*}
where the state vector $\vtheta$ contains the airlight color $\mA$ and the transmission map $\vt$, with $\odot$ defining the elementwise multiplication. Note that $\calH_{\vtheta}$ is a very good model for $\calG$, but $\calH_{\vtheta}\not= \calG$.

\begin{center}
	\includegraphics[width=\linewidth]{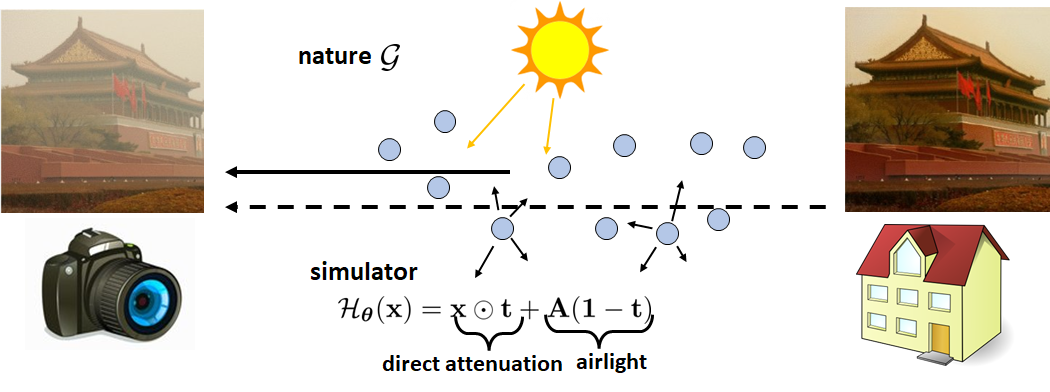}
\end{center}

}

\textbf{Simulator}. Since $\calG$ is unknown, we need to approximate it using a model $\calH_{\vtheta}$. I call it a simulator. A simulator, by construction, reproduces \emph{part} of the nature. Therefore, the output of the simulator is not $\vy$ but
\begin{equation}
	\widehat{\vy} = \calH_{\vtheta}(\vx).
\end{equation}
I emphasize that $\calH_{\vtheta}$ is a \emph{parameterization} of nature. The parameterization is a dimensionality reduction of the operator $\calG$ from an infinite-dimensional space to a finite (and often low) dimensional space. The choice of the parametric model is often based on physics, but there are also man-made parameterizations. The following is an example.

\boxedeg{

\vspace{2ex}

\textbf{Example 3}. (Environment parameterization) In \cite{Gnanasambandam_2021_ICCV}, Gnanasambandam et al. proposed the idea of optically perturbing the appearance of an object by using structured illumination. The concept was that given a ground truth image $\vx$ associated with a class label $\ell$, an optical system $\calH_{\vtheta}$ can be built such that the distorted image $\widehat{\vy} = \calH_{\vtheta}(\vx)$ will be misclassified as label $\ell'$. The optical system was implemented using a projector-camera setup, where the projector illuminates the real 3D object using a calculated pattern. This provides a mechanism to test the resilience of the classifier to manipulated attacks.

\begin{center}
\includegraphics[width=\linewidth]{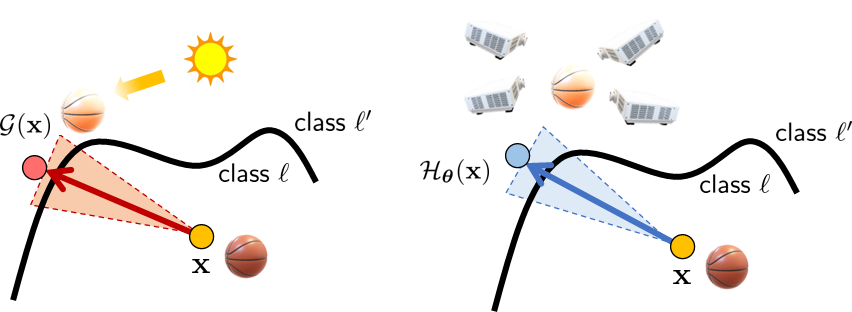}
\end{center}

What is relevant to CIF is that the same optical set up can be used to analyze the robustness against natural perturbations, such as shadow or overcast. These natural perturbations are $\calG$, of which the exact forms are unknown to us. Some people use graphics rendering (e.g., for flight simulator) to synthesize the scenes. With the projector-camera system, we can use the multiple projectors to \emph{parameterize} nature by using a finite number of tunable knobs to reproduce outdoor environment in a controlled setting, hence providing an approximation $\calH_{\vtheta} \approx \calG$.
}

\begin{figure}[h]
	\centering
	\includegraphics[width=\linewidth]{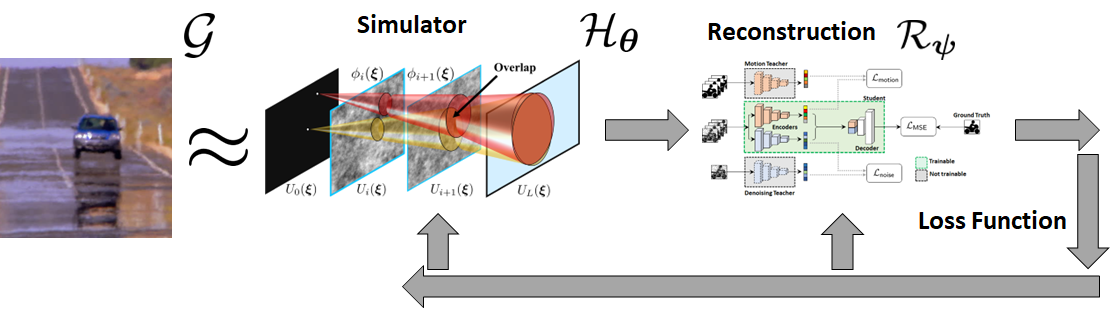}
	\caption{In Computational Image Formation (CIF), the true image formation process is determined by nature $\calG$. The design task here is to select and optimize a simulator $\calH_{\vtheta}$ such that it does not only achieve a desirable modeling accuracy, but more importantly maximize the reconstruction quality.}
	\label{fig: Ch0 computational imaging concept 2}
\end{figure}

The presence of nature and a simulator changes the problem scope from designing a hardware camera to designing a numerical simulator as shown in \fref{fig: Ch0 computational imaging concept 2}. Because of its middle-person role between nature and algorithm, the selection and optimization of the simulator is significantly more complicated than configuring a camera such as deciding which coded aperture pattern to use. An analogy here is that selecting a simulator is like selecting a neural network, whereas configuring a camera is like learning the weight of the neural network.

\subsection{Four Attributes of a CIF Simulator}
In what follows, I will describe the four attributes of the simulator in CIF.

\textbf{Attribute 1: Accurate}. The first attribute of a CIF simulator $\calH_{\vtheta}$ is that it is accurate enough to approximate the true image formation process $\calG$. The form of $\calH_{\vtheta}$ can be anything. It can be a one-line equation, an algorithmic procedure, or even a neural network. The performance of $\calH_{\vtheta}$ is measured in terms of how good the simulated image is when compared to the true image. Mathematically, we define the loss as
\begin{equation}
\calE_{\text{sim}}(\calH_{\vtheta}) \bydef
\underset{\text{simulator loss}}{\underbrace{\E_{\vx} \Big[ \text{SimLoss}(\calH_{\vtheta}(\vx), \calG(\vx))\Big]}}.
\label{eq: loss 1: simulator}
\end{equation}
The equation says that for a given image $\vx$, the simulated image $\calH_{\vtheta}(\vx)$ needs to be sufficiently similar to $\calG(\vx)$. The goodness of $\calH_{\vtheta}$ is measured according to the loss function $\text{SimLoss}$. For example, in atmospheric turbulence, SimLoss can be the pixelwise distance between the simulated long/short exposure function and the measured (real) long/short exposure function. It can also be the differential tilt statistics or the $z$-tilt statistics between the simulation and the theoretically derived statistical average. Or, SimLoss can be measured using target patterns seen through a heat chamber. There are also instrumentations such as the scintillometer to measure the turbulence profile. In any case, SimLoss is an application specific metric(s).

\textbf{Attribute 2: Fast}. The second attribute of a CIF simulator is that it needs to be fast so that we can integrate it as a part of the reconstruction algorithm. But speed depends on which computing platform we use. (See discussions in the next paragraph.) A slightly better way is to measure the complexity of the model, with the following notation
\begin{equation}
\calE_{\text{complexity}}(\calH_{\vtheta})
\label{eq: loss 2: speed}
\end{equation}
The definition of the complexity can take various forms. If $\calH_{\vtheta}$ is implemented using a deep neural network, then the complexity can be measured in terms of the number of hidden layers, width of hidden layers, number of parameters, number of filters, size of the filters, etc. The complexity of $\calH_{\vtheta}$ can also be viewed at two levels: (1) What is the intrinsic capacity of the model and (2) What is the effective model capacity given a specific choice of $\vtheta$.

Model complexity is usually linked to  speed (aka runtime), but some caveats should be noted. For example, 2D convolutions today can be efficiently implemented because of the dedicated hardware architectures on graphics processing units. In contrast, depth-wise 3D convolutions are substantially slower even if the number of parameters is the same as a 2D convolution because of the lack of specialized hardware. While this problem will likely be solved in the near future, the complexity is not translated to runtime directly.

\textbf{Attribute 3: Benefits Reconstruction}. The third attribute of a CIF simulator is that it needs to benefit the reconstruction performance because in CIF, the ultimate goal is to maximize the final image produced by the entire system.

The reconstruction loss is abstractly defined as
\begin{align}
\calE_{\text{recon}}(\calH_{\vtheta}) \bydef
\min_{\calR_{\vpsi}} \;
\underset{\text{reconstruction loss}}{\underbrace{
\E_{\vx} \Big[ \text{ReconLoss}\Big( \calR_{\vpsi}( \vy, \; \calH_{\vtheta} ), \vx \Big) \Big]
}}.
\label{eq: loss 3: reconstruction}
\end{align}
The reconstruction loss in \eref{eq: loss 3: reconstruction} is reminiscent to \eref{eq: comp camera 1}, but the goals are very different. In \eref{eq: comp camera 1}, the camera $\calH_{\vtheta}$ is already chosen except for its parameter $\vtheta$. Thus, the joint minimization in \eref{eq: comp camera 1} is to simultaneously optimize for the reconstruction (realized via a neural network) and the state vector $\vtheta$ of the camera. In contrast, the minimization problem in \eref{eq: loss 3: reconstruction} says that we do not even know which simulator to use. Using atmospheric turbulence as an example (see Section III), the decision could be the choice between two completely different models, e.g., the pixel-jitter model or the deformable model. Therefore, in CIF simulator selection, the difficulty is how to \emph{build} a simulator that fulfills the criteria instead of adjusting parameters of some simple equations.

\textbf{Attribute 4: Differentiable}. The fourth attribute of a simulator is that it needs to be differentiable. The metric for differentiability can be a simple indicator function:
\begin{align}
\calE_{\text{diff}}(\calH_{\vtheta}) \bydef
\begin{cases}
0,      & \text{if $\calH_{\vtheta}$ is differentiable},\\
\infty, & \text{if $\calH_{\vtheta}$ is not differentiable}.
\end{cases}
\label{eq: loss 4: differentiability}
\end{align}

Why do we need differentiability? The reason is that most, if not all, image reconstruction algorithms today are based on deep neural networks. For the simulator to be part of the reconstruction framework, it is necessary that the gradient of the loss function can be backpropagated to the input. Therefore, being able to take the gradient of the simulator with respect to its model parameters becomes essential.

Enforcing differentiability can be realized in multiple ways. The easiest way, of course, is to build a simulator using a deep neural network. However, speaking from experience, physics-based simulators often offer significantly better guarantees of the theoretical properties and are more interpretable. Yet, physics-based simulators are often complicated. Enforcing differentiability is not a trivial task. For example, iterative procedures such as Newton's method, which is widely used to find the intersection of a ray and the lens surface, needs to be replaced by another method that is non-iterative before we can make it differentiable.

\subsection{Simulator Selection and Parameter Optimization}
After stating the four attributes of the simulator, I would like to elaborate on the simulator selection problem and the parameter optimization problem. The two problems are different. The former is more about \emph{building} a simulator that fulfills the criteria. The latter assumes that the candidate simulator has already been decided, but during the inference process we would like to update the state vector to better match with the observation. The simple analogy here is that simulator selection picks the number of partitions in a spatially varying blur system (i.e., how many pixels to share one blur kernel), whereas the parameter update assumes that the partition is already fixed but we need to estimate the individual blur kernels.

\textbf{Simulator Selection}. The simulator selection problem can be formulated as a constrained optimization.
\begin{align}
&\argmin{\calH_{\vtheta}}    &&\calE_{\text{recon}}(\calH_{\vtheta}) + \calE_{\text{diff}}(\calH_{\vtheta}) \notag\\
&\text{subject to }          &&\calE_{\text{complexity}}(\calH_{\vtheta}) \le \tau_{\text{complexity}},\\
&                            &&\calE_{\text{sim}}(\calH_{\vtheta}) \le \tau_{\text{sim}}\notag
\end{align}
The objective function of the problem is the reconstruction loss $\calE_{\text{recon}}(\calH_{\vtheta})$ plus the indicator function $\calE_{\text{diff}}(\calH_{\vtheta})$. Since by construction the simulator needs to be differentiable, the indicator function will serve the purpose of only allowing differentiable simulators to be considered. The main objective is therefore the reconstruction loss.

There are two constraints in the optimization. The complexity constraint limits how complex the simulator can be. The simulator loss constraint controls the match between the simulator $\calH_{\vtheta}$ and the ground truth model $\calG$. The reason why these two criteria are listed as constraints instead of the objective is that in practice, the simulator designer would seldom navigate the four criteria simultaneously when choosing the simulator. It is far more common to develop a simulator with a certain level of complexity and simulation accuracy in mind, while leaving some design freedom to the reconstruction algorithm.

I should stress that while the simulator selection problem appears like an ordinary constrained optimization, the problem is never solved using any gradient based methods. In contrast, the problem is more often solved manually based on experience and creativity. So it is more of a piece of art.

\textbf{Parameter Optimization}. To motivate the parameter update discussion, it will be best to first consider an example.
\boxedeg{

\vspace{2ex}

\textbf{Example 4}. (Blind deconvolution) In the classical blind image deconvolution problem, we are often interested in the alternating minimization
\begin{align*}
\vh^{k+1} &= \argmin{\vh} \; \|\vh \circledast \vx^k - \vy\|^2     + \text{prior}(\vh),\\
\vx^{k+1} &= \argmin{\vx} \; \|\vh^{k+1} \circledast \vx - \vy\|^2 + \text{prior}(\vx),
\end{align*}
where $\text{prior}(\cdot)$ is a generic notation representation the prior model of the respective variable.

The first equation, put into the CIF framework, can be rewritten as
\begin{equation*}
\vtheta^{k+1} = \argmin{\vtheta} \;\; \Big\{\text{SimLoss}\left( \calH_{\vtheta}(\vx^k), \calG(\vx) \right)\Big\}.
\end{equation*}
Here, $\calG$ is the true image formation process given by nature, and $\calH_{\vtheta}(\vx) = \vh \circledast \vx + \vn$ is the approximation. The second equation can be written as
\begin{equation*}
\vx^{k+1} = \calR_{\vpsi}(\vy, \calH_{\vtheta^k}),
\end{equation*}
which is a generic form of the image reconstruction method.

\begin{center}
\includegraphics[width=\linewidth]{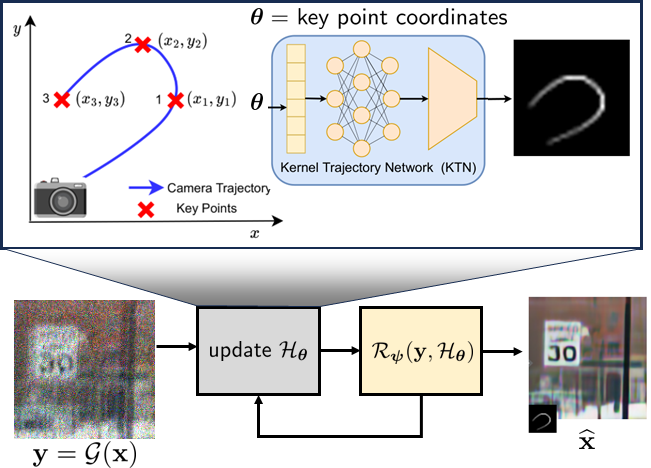}
\end{center}

To be slightly more precise about the state vector $\vtheta$, one can think of it as the parametric description of the forward model $\calH_{\vtheta}$. For example, in \cite{Sanghvi_2023_CVPR}, key points are used to characterize the motion where a continuous curve is drawn to connect the key points to generate a motion blur kernel. The key points are coordinates in the 2D space, and the state vector $\vtheta$ is the collection of all key points.
}

The above example illustrates the typical procedure of updating the simulator's parameters and estimating the latent image. These two steps can be summarized by the following pair of iterative updates:
\begin{align}
\vx^{k+1}     &= \calR_{\vpsi}( \vy, \; \calH_{\vtheta^{k}} ) \\
\vtheta^{k+1} &= \argmin{\vtheta}\;\; \text{SimLoss}(\calH_{\vtheta}(\vx^{k+1}), \calG(\vx))
\end{align}
As I mentioned earlier, this pair of equations are performed \emph{after} the simulator is selected.

\subsection{Discussions}
\textbf{Ideal $\calG$?} At this point, it should be clear why a perfect simulator $\calH_{\vtheta} = \calG$ may not be a good simulator in CIF. If $\calG$ is intrinsically complicated, then matching it would require a complex $\calH_{\vtheta}$. This will violate Attribute 2 (simple and fast), and likely Attribute 4 (differentiable). There is also a possibility that Attribute 3 (improve reconstruction) will be poor because a complex $\calH_{\vtheta}$ will often lead to an ill-posed inverse problem. The reconstruction model $\calR_{\vpsi}$ has to be even more complex in order to invert the effect of $\calG$. This means that we have a good forward model, but we cannot solve the inverse problem.

\textbf{Compared to a physics simulator}. Compared to a pure physics simulator, a CIF simulator has an extra emphasis on speed, reconstruction, differentiability and compatibility with the reconstruction algorithm. This leads to the summary in Table~\ref{tab: simulator comparison}.

\begin{table}[h]
\centering
\caption{Comparison between a physics simulator and CIF simulator}
\label{tab: simulator comparison}
\scalebox{0.8}{
\begin{tabular}{p{1.6cm}p{3.8cm}p{3.8cm}}
\hline
            & \textbf{Physics simulator}     & \textbf{CIF simulator}\\
\hline
\hline
Goal        & Describe nature       & Help image reconstruction \\
\hline
Accurate?    & As much as possible because we need it to reproduce nature.
            & No need to be absolutely accurate. As long as it meets a certain level, it is okay.\\
\hline
Fast?       & Fast is certainly welcome, but slow is also okay if it is justified by application.
            & Need to be very fast, for synthesizing datasets and integrating with inverse algorithms.\\
\hline
Benefits Reconstruction? & Irrelevant because the simulator's goal is to reproduce nature.
               & Simulator in the reconstruction loop is the key of CIF.\\
\hline
Differentiable? & Irrelevant. No need & Needed, especially if we want to co-optimize reconstruction and simulator.\\
\hline
\end{tabular}
}
\end{table}

\textbf{Compared to computational camera}. How is CIF different from a computational camera? In a computational camera, the optimization variable is the hardware camera. For example, we adjust the lens parameter, or the mask in coded aperture, or the pattern of the exposure multiplexing scheme. CIF \emph{can} in principle be a superset because these hardware elements can always be described by mathematical models. However, the bigger difference (and also the problem context) is that CIF is more relevant to situations where $\calG$ is not easily parameterized by simple equations. Weather is the best example to think about. In those cases, \emph{choosing} which $\calH_{\vtheta}$ is more relevant than optimizing for $\vtheta$. In short, CIF is not only about the co-optimization of the image formation elements, but also about the selection of the image formation model.

\textbf{Compared to model selection literature}. The subject of simulator selection is many ways similar to the classical topic on model selection \cite{Ding_2018_SPM}. The core difference, however, is that in model selection, the models are more or less simple equations, for example, the $\ell_1$-norm (LASSO) or the $\ell_2$-norm (ridge). In CIF, the selection of simulator can be drastically more difficult. Often times, there may not be any readily available simulators to ``choose from''. The designer needs to develop a simulator to match the complexity and simulation loss, while maximizing the image reconstruction performance. This is \emph{not} a simple gradient search problem, but more of a design problem.

Because of the different goals and different problem settings, the methodologies used in model selection largely do not apply. For example, tools such as the Akaike information criterion (AIC) or the Bayesian information criterion (BIC) are not applicable because (i) even though there is an underlying probability distribution of the true data, we never know what it is, (ii) most simulators are complex numerical procedures. Analyzing them and calculating in the context of AIC and BIC is infeasible, (iii) there exist domain-specific alternative assessment techniques that are much more reliable and direct than AIC and BIC.

\begin{figure}[h]
\centering
\includegraphics[width=\linewidth]{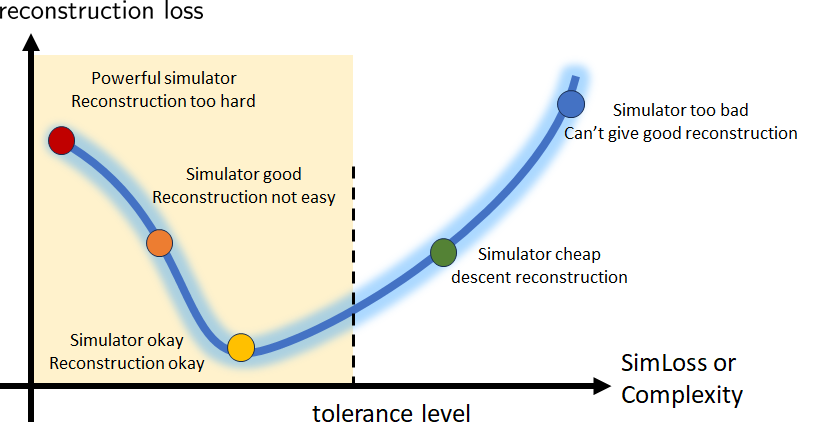}
\caption{A conceptual illustration of the trade off between the performance of the simulator and the image reconstruction.}
\label{fig: trade off}
\end{figure}

\textbf{Trade-off}. Because of the conflicting goals in the simulator selection problem, it would be useful to see the tension via the illustration in \fref{fig: trade off}. In this figure, the $y$-axis defines the reconstruction loss whereas the $x$-axis defines the simulator loss (or complexity). The question of CIF is about which point on the trade-off curve to pick.

On the far right of the figure, the simulator is too simple to be of any usage. Even though the reconstruction is easy, the reconstructed image will not be good. Moving slightly to the left, the simulator is improved and the reconstruction is also improved thanks to a more meaningful simulator. The benefit can come from the data synthesis (i.e. to generate a good training dataset), as well as in-the-loop inverse algorithm. The optimal operating point is somewhere in the middle, where the simulator is reasonably accurate and the reconstruction is also descent. As the simulator moves towards the left hand side, the reconstruction becomes more difficult. Not only do we have a harder time to synthesize enough data because the simulation can be time-consuming, but we also have a harder time to put the simulator in the reconstruction loop. As the simulator moves to the far left, the simulator is extremely good but it offers little to no help to the reconstruction. Thus, the performance can be very bad. In the figure, I also marked a colored region with a cutoff $\tau$. Any simulator that gives a performance better than $\tau$ will be inside the region. Therefore, among all the possible simulators, only those that meet the simulator criteria can be adopted for reconstruction.

\section{Case Study: Imaging through Turbulence}
Now that we have seen what CIF means, it is time to look at a concrete example. I will use imaging through atmospheric turbulence as an illustration because of the richness of the literature. To give you an idea of the accuracy of the model, I will use the number of stars $\star$ to indicate how accurate the simulator is. A simulator with one ($\star$) means that it is cheap but inaccurate, whereas a simulator with ($\star\star\star\star\star$) means that it is extremely accurate. The number of stars is partially objective because we can measure the quality, but it is also partially subjective because much of these are based on my experience.

\subsection{Wave Propagation Physics}
Our atmosphere is a complicated medium because of its turbulent nature. Factors such as temperature, wind velocity, humidity, and other weather conditions can all affect the strength of turbulence. As light propagates, the random index of refraction will cause phase delays along the optical paths, hence leading to pixel displacements and blurs \cite{Tatarski_1967_a, Roggemann_1996_a}. The exact image formation for one specific turbulent instance cannot be determined because it is stochastic.

\vspace{1ex}
\textbf{Split-step} (``Gold standard'') ($\star\star\star\star\star$) Let me start by discussing the most accurate model so that we have a reference point. In physics, the most accurate model for atmospheric turbulence, to date, is the split-step propagation \cite{Schmidt_2010_a}. The model says that we can partition the propagation path into a sequence of segments where each segment has a random phase screen determining the phase distortion. The propagation is performed by integrating the electromagnetic field $\vu$ of which the magnitude produces the image $\vx = |\vu|^2$.

\begin{figure}[h]
\centering
\includegraphics[width=\linewidth]{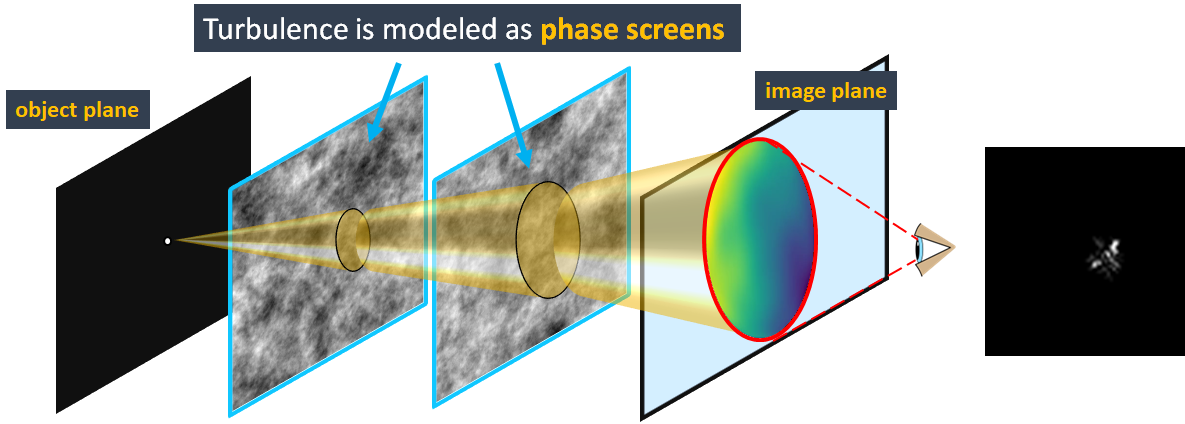}
\caption{The split-step propagation model $\calH_{\vtheta}^{\text{split}}$ is considered as the ``gold standard'' in modeling atmospheric turbulence. While the model is accurate, its numerical wave propagation equation makes the implementation extremely difficult and very slow. Thus, it is a good simulator for mimicking nature, but not so much for image restoration.}
\label{fig: split}
\end{figure}

Without diving into the technical details, we can think of the split-step propagation as a sequence of operations along the light propagation path. At the $i$th segment, the electromagnetic field propagates according to the equation
\begin{equation}
\vu_{i+1} = \text{Fresnel}( \underset{\text{$i$th field} }{\underbrace{\vu_i}} \;\odot\; \underset{\text{$i$th phase screen}}{\underbrace{e^{j \vphi_i}}}),
\end{equation}
where $\odot$ denotes elementwise multiplication, $\text{Fresnel}(\cdot)$ denotes the Fresnel diffraction integral \cite{Goodman_2005_a, Goodman_2015_a}, and $\vphi_i$ denotes the $i$th phase screen drawn from the Kolmogorov power spectral density \cite{Kolmogorov_1941_a, Frisch_1995_a, Lane_1992_a}.\footnote{Here I am assuming a ``typical'' situation with a small to moderate Rytov number so that we can ignore the amplitude effect and the power spectral density roughly follows the $5/3$-power law predicted by Kolmogorov \cite{Bos_2015_a}.} The process is nonlinear because for every pixel in the object plane, the equation has to be evaluated repeatedly (which includes Fourier transform, phase cropping, and multiplication), as illustrated in \fref{fig: split}. For completeness, let me write down the equation
\begin{equation}
\widehat{\vy} = |\underset{\calH_{\vtheta}^{\text{split}}(\vx)}{\underbrace{\text{Fresnel}(\text{Kolmgrv}(\ldots(\text{Fresnel}(\text{Kolmgrv}(\vx))))}}|^2,
\end{equation}
where $\calH_{\vtheta}^{\text{split}}$ denotes the split-step model, with $\vtheta$ defining the random phase screens along the propagation path.

For decades, physicists have agreed that the split-step propagation is very reliable and accurate for simulating atmospheric turbulence. With a sufficient number of phase screens along the path, and assuming that the turbulence is at most moderately strong, it is safe to say that
\begin{equation}
\calG(\vx) \approx \calH_{\vtheta}^{\text{split}}(\vx).
\end{equation}
Various simulation packages have been built \cite{Leonard_2012_a, Potvin_2011_a, Hardie_2017_a, Roggemann_2012_a, Schwartzman_2017_a, Vorontsov_2005_a, Lachinova_2007_a, Roggemann_1995_a, Repasi_2008_a, Potvin_2020_a, Hardie_2017_b}, and they are widely used to study the optical communication, high energy beam propagation, and astronomical applications. There are also attempts to use these simulators for image processing tasks, but people generally find split-step too slow for any real-time applications. The bigger problem arises if we want to use $\calH_{\vtheta}^{\text{split}}$ for image restoration. Let's see why.

\subsection{Image Restoration}
Split-step propagation is never used as part of the image restoration pipeline. I will illustrate this using two reconstruction methods.

The first approach is the optimization approach. Using total variation as one of (many) examples, we have
\begin{equation}
\vxhat = \calR_{\vpsi}(\vy) = \argmin{\vx} \; \| \calH_{\vtheta}^{\text{split}}(\vx) - \vy  \|^2 + \lambda \|\vx\|_{\text{TV}}.
\end{equation}
But solving this problem can be hard because $\calH_{\vtheta}^{\text{split}}$ is complicated. If we perform variable-splitting techniques such as the ADMM algorithm, we largely have no way of computing the proximal map because the inverse (or regularized inverse) $(\calH_{\vtheta}^{\text{split}})^{-1}$ is not possible to obtain. Note that here I am not even talking about estimating the state vector $\vtheta$, which will add more difficulties to the problem.

Another approach is deep learning. The idea is to train a deep neural network $\calR_{\vpsi}$ that performs
\begin{equation}
	\vxhat = \calR_{\vpsi}(\vy,\calH_{\vtheta}^{\text{split}}).
\end{equation}
However, the problem is how to generate the clean-noisy pair for training. We can send the clean image through $\calH_{\vtheta}^{\text{split}}$, and repeat the process until we have created a dataset of training samples. This is in theory doable, but since the simulator is so slow, there is no way we can synthesize enough data \cite{Mao_2021_a}.

There is one more issue about deep learning. If we want to incorporate $\calH_{\vtheta}^{\text{split}}$ into the reconstruction neural network $\calR_{\vpsi}$, the simulator needs to be \emph{differentiable}. Since the split-step simulation is a serial chain of complex operations going back and forth with the Fourier transforms per pixel, back propagating the gradient of the loss function through the simulator is nearly impossible unless we introduce additional approximations.

\boxedeg{
\vspace{1ex}
CIF Trade-off for \textbf{Split-step}:
\begin{itemize}
\setlength\itemsep{0ex}
\item $\calE_{\text{sim}}$: Extremely small, especially if we use enough phase screens.
\item $\calE_{\text{complexity}}$: Huge.
\item $\calE_{\text{recon}}$: Unknown, because the only way split-step can be used is to synthesize training data. But even so, it is computationally infeasible because split-step is too slow.
\item $\calE_{\text{diff}}$: Infinity, because split-step needs to perform a sequential chain of Fourier transforms of phase functions per pixel.
\end{itemize}
}

\subsection{Lightweight Simulators}
Having discussed the ``gold standard'' split-step simulator, let me now switch gear to talk about something faster.

\begin{figure}[h]
\centering
\begin{tabular}{cc}
\includegraphics[width=0.45\linewidth]{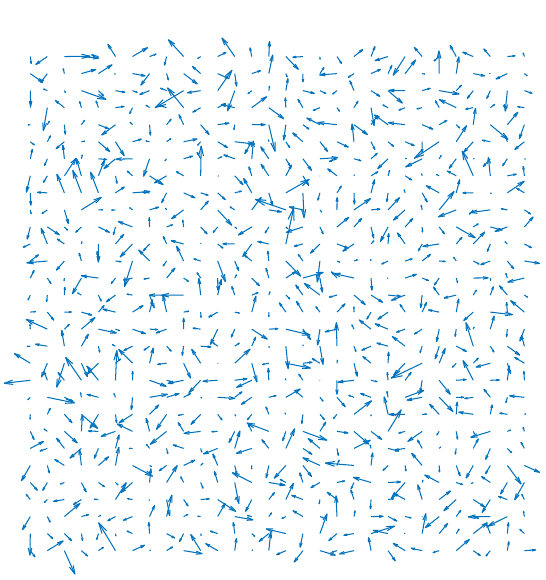}&
\includegraphics[width=0.45\linewidth]{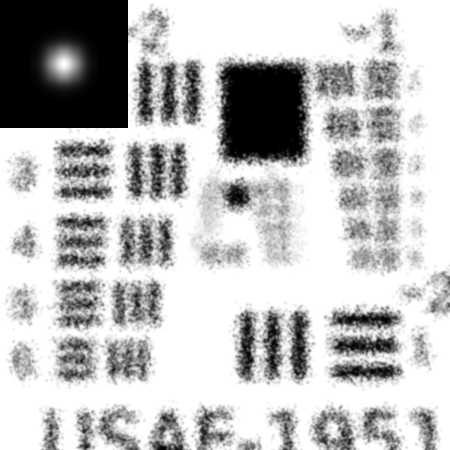}\\
(a) Random jitter & (b) Blur kernel and image
\end{tabular}
\caption{Random jitter and spatially invariant Gaussian blur. The simulated turbulence effect is quite far from the reality.}
\label{fig: lightweight 01}
\end{figure}

\textbf{Pixel Jitter Model} \cite{Leonard_2012_a}. ($\star$) The simplest simulator reported in the literature is the pixel shift + blur model. This model says that
\begin{equation}
	\widehat{\vy} = \underset{\calH_{\vtheta}^{\text{jitter}}(\vx)}{\underbrace{\vh \circledast ( \text{jitter} (\vx))}},
\end{equation}
where $\text{jitter}(\cdot)$ denotes the pixel jittering, where the displacement is a random vector with an i.i.d. distribution. The operator $\vh$ is a spatially-invariant Gaussian blur, and $\circledast$ is the convolution. \fref{fig: lightweight 01} shows an example. $\calH_{\vtheta}^{\text{jitter}}$ is very cheap and fast -- just draw a random i.i.d. motion vector field (pick any distribution, e.g., Gaussian), and convolve the image with an invariant blur kernel. The state vector $\vtheta$ contains the blur and the motion vector field.

Needless to say, inverting $\calH_{\vtheta}^{\text{jitter}}$ is easy, at least in the non-blind case where we know the random vector field and the blur: Just remove the blur using any off-the-shelf non-blind deconvolution algorithm, and then un-do the pixel displacement. We can plug it into a neural network. We can also plug it into a classical optimization so we can solve the proximal map. The only caveat (and a big caveat) is that $\calH_{\vtheta}^{\text{jitter}} \not\approx \calG$, not even close.

\boxedeg{
\vspace{1ex}
CIF Trade-off for \textbf{Pixel-Jitter Model}:
\begin{itemize}
\setlength\itemsep{0ex}
\item $\calE_{\text{sim}}$: Huge, because the simulator is too simple.
\item $\calE_{\text{complexity}}$: Extremely small.
\item $\calE_{\text{recon}}$: Poor, because the simulator is too far from reality although solving the reconstruction problem is easy. Data synthesized by the model will have a hard time generalizing to real turbulence.
\item $\calE_{\text{diff}}$: Zero, because both pixel jitter and blurring steps are differentiable.
\end{itemize}
}

\begin{figure}[h]
\vspace{-2ex}
\centering
\begin{tabular}{cc}
\includegraphics[width=0.45\linewidth]{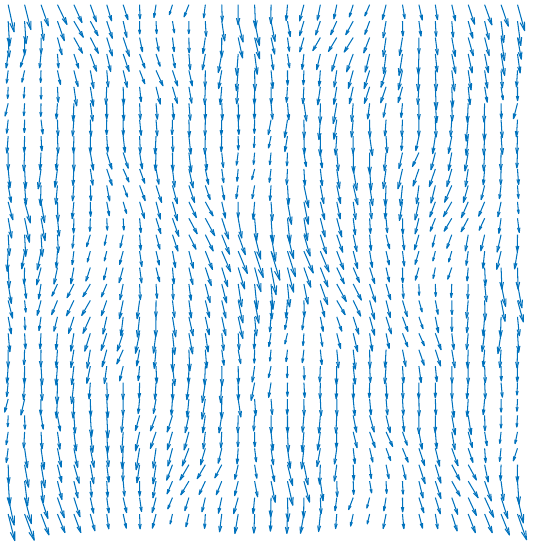}&
\includegraphics[width=0.45\linewidth]{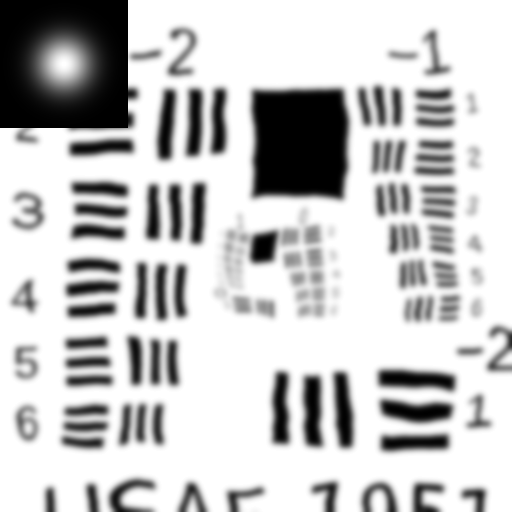}\\
(a) Deformable field & (b) Blur kernel and image
\end{tabular}
\caption{Deformable field and spatially invariant Gaussian blur. The simulated turbulence effect is better than the random jitter, but still not accurate enough to match the real turbulence.}
\label{fig: lightweight 02}
\end{figure}

\vspace{1ex}
\textbf{Deformable Model} \cite{Shimizu_2008_a, Gilles_2012_a, Gilles_2016_a, Lou_2013_a, Milanfar_2013_a}. ($\star\star$) A slightly better version of the simulator is the deformable grid model. Instead of assuming an i.i.d. random displacement field, we select a small set of anchor points in the image grid and perform a non-rigid deformation. The overall equation is
\begin{equation}
	\widehat{\vy} = \underset{\calH_{\vtheta}^{\text{deform}}(\vx)}{\underbrace{\vh \circledast ( \text{deform} (\vx))}},
\end{equation}
The benefit of the deformable model is that it is physically more meaningful because the pixel displacement caused by turbulence is spatially correlated. The deformable model provides a way to enforce the spatial correlation. \fref{fig: lightweight 02} illustrates an example, where we can see that the deformable grid is more structured. The sampling process is reasonably easy because we only need to identify a few anchor points.

The deformable grid facilitates the estimation of the pixel displacement because the number of anchor points is finite. The drawback, though, is that the blur remains spatially invariant. It is still valid for isoplanatic turbulence, but definitely not true for anisoplanatic turbulence where the blur must be spatially varying.

\boxedeg{
\vspace{1ex}
CIF Trade-off for \textbf{Deformable Model}:
\begin{itemize}
\setlength\itemsep{0ex}
\item $\calE_{\text{sim}}$: Better than the pixel-jitter model.
\item $\calE_{\text{complexity}}$: Small. Comparable to pixel-jitter.
\item $\calE_{\text{recon}}$: Not great but better than pixel-jitter. The simulator is still far from reality although solving the reconstruction problem is easy. Data synthesized by the model does not generalize to real turbulence.
\item $\calE_{\text{diff}}$: Zero. The model is differentiable.
\end{itemize}
}

\begin{figure}[h]
\centering
\includegraphics[width=\linewidth]{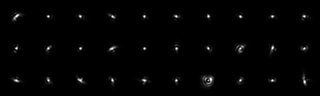}
\caption{Spatially varying blur kernels used in $\calH_{\vtheta}^{\text{varying}}$. The kernels are generated using the phase-over-aperture model. Note that the shape of the blur changes from one location to another.}
\label{fig: lightweight 03}
\end{figure}

\vspace{1ex}
\textbf{Varying Blur Model}. ($\star\star\star$ Incomplete) To model anisoplanatic turbulence, people propose to adopt a spatially varying blur. Keeping the deformable grid idea (which can be replaced by other pixel displacement models), the output image is defined in a per-pixel basis:
\begin{equation}
\widehat{\vy}_i = \underset{\calH_{\vtheta}^{\text{varying}}(\vx)}{\underbrace{\vh_i \circledast ( \text{deform} (\vx))}},
\end{equation}
where $i$ denotes the $i$th pixel of the image $\widehat{\vy}$ and $\vh_i$ denotes the blur for the $i$th pixel. Because of the per-pixel blur, this model can now describe a wider set of turbulence effects. However, the generation of these per-pixel blurs remains unclear (that's why I mark it as ``incomplete''). In the example blur kernels I show in \fref{fig: lightweight 03}, I use the phase-over-aperture model (to be discussed in the next subsection) to generate the spatially varying blur. These blurs are then integrated into this model $\calH_{\vtheta}^{\text{varying}}$ to perform the distortion.

The inverse problem associated with $\calH_{\vtheta}^{\text{varying}}$ is not easy, because spatially varying blurs do not have simple inverses using standard tools such as the Fourier transform. Therefore, additional approximations are needed to reduce complexity.

\boxedeg{
\vspace{1ex}
CIF Trade-off for \textbf{Varying Blur Model}:
\begin{itemize}
\setlength\itemsep{0ex}
\item $\calE_{\text{sim}}$: Undetermined, because it is unclear how the per-pixel blur kernels are defined while satisfying the spatial correlation of the blur.
\item $\calE_{\text{complexity}}$: Undetermined, because it depends on how the kernels are generated. Assuming that the covariance matrix is given, then sampling from it is not too complicated.
\item $\calE_{\text{recon}}$: Small if the spatial correlation of the blur kernels are satisfied.
\item $\calE_{\text{diff}}$: Zero. The model is differentiable if the blur kernels are predetermined and stored.
\end{itemize}
}

\vspace{1ex}
\textbf{Flipped Model} \cite{Lou_2013_a, Lau_2019_a, Lau_2021_a}. ($\star\star\star$) Because of the difficulty of inverting the spatially varying blur, people propose a work-around solution by observing that turbulence-distorted images exhibit a lucky effect. From time to time, and from pixel to pixel, there are instants where the turbulence distortion is weak. The lucky pixels / patches existing in the raw video input are blur-free and displacement-free. Therefore, if we can identify these lucky patches, then we can mitigate majority of the turbulence effects, leaving only the diffraction limited blur. This two step procedure can be summarized as
\begin{align}
\widehat{\vx} =
\underset{\text{reconstruction}}{\underbrace{
\text{deblur} \left( \text{lucky} (\vy) \right)}}.
\end{align}
There are many variations of this approach, such as \cite{Shimizu_2008_a, Mao_2020_a, Anantrasirichai_2013_a, Anantrasirichai_2018_a, Milanfar_2013_a}.

The two-step procedure inspired people to consider an alternative model where we flip the order of blur and pixel displacement in the simulator. This gives us the flipped model:
\begin{equation}
\widehat{\vy}_i = \underset{\calH_{\vtheta}^{\text{flipped}}(\vx)}{\underbrace{\text{deform} (\vh_i \circledast \vx)}},
\end{equation}
which according to \cite{Chan_2022_a}, is called the blur-tilt model. The blur-tilt model is not the same as the tilt-blur model because it can be proved that wave propagation follows tilt-blur but not blur-tilt. However, the blur-tilt model is easier from the angle of inverse problem because inverting the deformation can be realized by lucky imaging algorithms.

\begin{figure}[h]
\centering
\begin{tabular}{cc}
\includegraphics[width=0.45\linewidth]{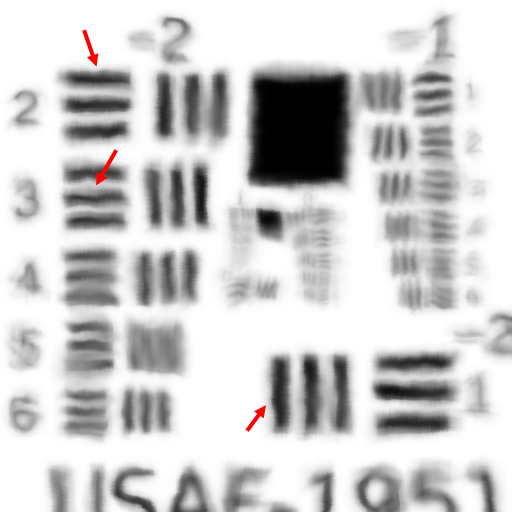}&
\includegraphics[width=0.45\linewidth]{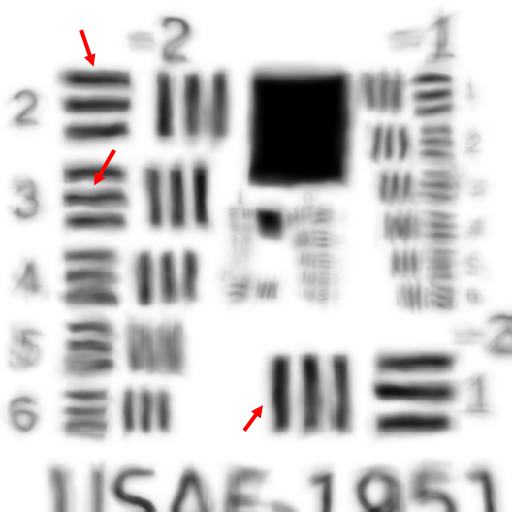}\\
(a) blur-tilt $\calH_{\vtheta}^{\text{flipped}}$ & (b) tilt-blur $\calH_{\vtheta}^{\text{varying}}$
\end{tabular}
\caption{Two models: Tilt-then-blur $\calH_{\vtheta}^{\text{varying}}$ and blur-then-tilt $\calH_{\vtheta}^{\text{flipped}}$. They generate similar result for most regions but not for edge pixels. As shown in \cite{Chan_2022_a}, $\calH_{\vtheta}^{\text{flipped}}$ has a tendency to ``destroy'' the blurs}
\label{fig: lightweight 04}
\end{figure}

\boxedeg{
\vspace{1ex}
CIF Trade-off for \textbf{Flipped Model}:
\begin{itemize}
\setlength\itemsep{0ex}
\item $\calE_{\text{sim}}$: Worse than the varying model because we can theoretically prove that the order of blur and tilt should be tilt-then-blur.
\item $\calE_{\text{complexity}}$: Same complexity as the varying model because it is the same composition of blur and tilt.
\item $\calE_{\text{recon}}$: Comparable to the varying blur model but the flipped model is easier to implement and hence it facilitates image reconstruction.
\item $\calE_{\text{diff}}$: Zero. The model is differentiable.
\end{itemize}
}

\textbf{Summary}. As you can see, the consideration of which $\calH_{\vtheta}$ to use is not a simple question of matching physics. If it was, then any of the cheap models would have no value. But the fact that they are used in the literature and sometimes generates satisfactory results speak of their validity. They are not justified from the angle of physics, but the angle of image reconstruction.

\subsection{Differentiable Simulator}
Over the past few years, there is an increasing amount of efforts to build faster and more accurate turbulence simulators with the goal of maximizing the image reconstruction performance. One of these efforts is a series of work of random sampling in the \emph{phase} domain. In this subsection I briefly comment on their basic principles, and highlight a few key attributes, in particular differentiability and scalability.

\textbf{Phase-over-aperture} \cite{Chimitt_2020_a} ($\star$ $\star$ $\star$ $\star$, not differentiable)  One of the biggest hurdles of the spatially varying blur model $\calH_{\vtheta}^{\text{varying}}$ is that there is no simple way of constructing the spatially varying blur kernels while satisfying the turbulence physics. In addition, the pixel displacement motion field is also not a simple deformable grid. These two limitations require a new method to model the turbulence.

\begin{figure}[h]
\centering
\includegraphics[width=\linewidth]{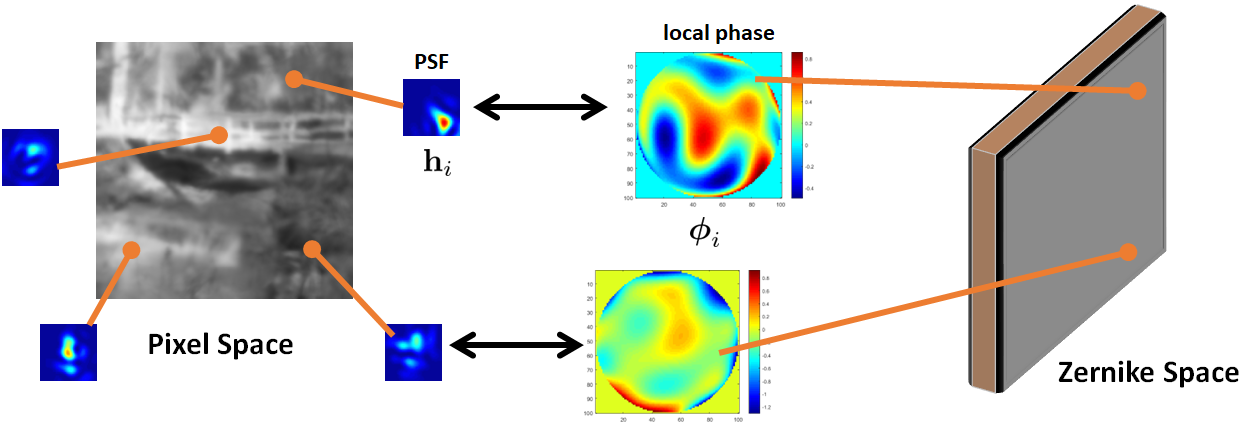}
\caption{Phase-over-aperture model assumes that every point spread function has a corresponding phase function. Thus, as long as we can generate the phase functions efficiently, we can construct the point spread function \emph{without} using numerical propagation techniques.}
\label{fig: aperture}
\end{figure}

The idea behind the phase-over-aperture model, proposed in \cite{Chimitt_2020_a}, was to recognize that the overall distortion (blur+displacement) is constructed from the Fourier magnitude square of the phase:
\begin{equation}
\vh_i = |\text{Fourier}(e^{j \vphi_i})|^2,
\end{equation}
where $\vphi_i$ is the phase function of the $i$th pixel. The displacement motion vector can be obtained by identifying the centroid of $\vh_i$. Any displacement from the centroid will be marked as the pixel displacement vector. Therefore, for every pixel $i$, there is a one-to-one mapping from the spatial domain to the phase domain, as illustrated in \fref{fig: aperture}.

For any $i$, the phase function $\vphi_i$ has a basis representation using the Zernike polynomials:
\begin{equation}
\underset{\text{phase function}}{\underbrace{\vphi_i}} = \sum_{j=1}^M \quad \underset{\text{coefficients}}{\underbrace{ \quad \alpha_{i,j} \quad}} \;\; \cdot \;\; \underset{\text{Zernike polynomials}}{\underbrace{\quad \mZ_j \quad }}.
\end{equation}
Since the Zernike polynomials are known functions, the blur kernel $\vh_i$ is constructed as soon as the coefficients $\valpha = \{\alpha_{i,j}\,|\, i = 1,\ldots,N, j = 1,\ldots,M\}$ become available.

The construction of the coefficient $\alpha_{i,j}$ can be done by sampling it from an enormous covariance matrix if we follow Tatarskii's assumption that the underlying random process is Gaussian \cite{Tatarski_1967_a}. In this case, by following the tedious derivations shown in \cite{Chimitt_2020_a}, the coefficients can be obtained by
\begin{equation}
\valpha = \text{Sampling from }( \E[\valpha\valpha^T] ),
\end{equation}
where $\E[\valpha\valpha^T]$ denotes the correlation matrix. The correlation matrix can be derived from the turbulence statistics where there are formulae to employ. Depending on how much turbulence statistics is utilized, $\valpha$ can maintain some degree of spatial correlations from one pixel in the scene to another pixel, and from one Zernike basis to another Zernike basis. If the turbulence condition is changed, then the correlation matrix will change and so the random coefficients change too.

\begin{figure}[h]
\vspace{-2ex}
\centering
\begin{tabular}{cc}
\includegraphics[width=0.45\linewidth]{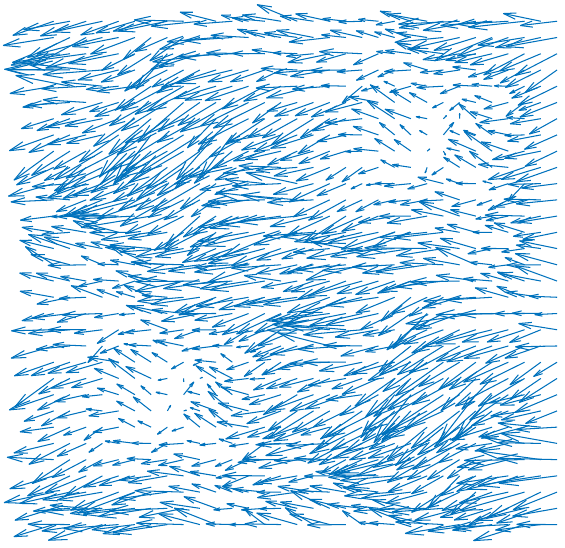}&
\includegraphics[width=0.45\linewidth]{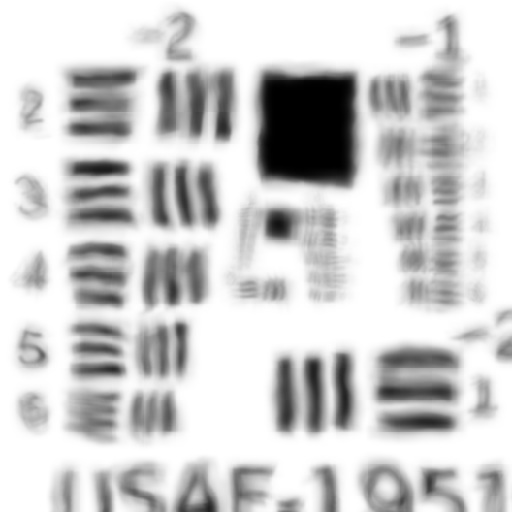}\\
(a) Displacement vector & (b) Resulting image
\end{tabular}
\caption{Phase-over-aperture \cite{Chimitt_2020_a}. (a) The displacement motion field, and (b) The resulting image. Note that the spatially varying blur in \fref{fig: lightweight 04} is generated using this phase-over-aperture model.}
\label{fig: phase over aperture example}
\end{figure}

\fref{fig: phase over aperture example} shows a random motion field generated by the phase-over-aperture model, as well as the resulting image.

The bottomline of the simulator is that it turns the split-step propagation equation into a random sampling method. This will give us a model
\begin{equation}
\widehat{\vy}_i =
\underset{\calH_{\vtheta}^{\text{phase}}(\vx_i)}{\underbrace{
\vh_i \circledast \left[ \vt_i \circ \vx_i \right]}},
\end{equation}
where $\vt_i$ is the pixel displacement vector that can be extracted from the first two Zernike polynomial basis. The operator $\circ$ denotes the pixel displacement operation.

\boxedeg{
\vspace{1ex}
CIF Trade-off for \textbf{Phase-Over-Aperture Model}:
\begin{itemize}
\setlength\itemsep{0ex}
\item $\calE_{\text{sim}}$: Very close to $\calH_{\vtheta}^{\text{split}}$ except for extreme turbulence levels beyond the Kolmogorov regime.
\item $\calE_{\text{complexity}}$: More complex than the deformable model, but still much simpler than the split-step propagation method.
\item $\calE_{\text{recon}}$: Small. However, the usage of $\calH_{\vtheta}^{\text{phase}}$ is limited to generating training datasets. No simulator in the loop because $\calH_{\vtheta}^{\text{phase}}$ is not differentiable.
\item $\calE_{\text{diff}}$: Infinity. The model is not differentiable.
\end{itemize}
}

\vspace{1ex}
\textbf{Phase to Space} ($\star$ $\star$ $\star$ $\star$, differentiable) \cite{Mao_2021_a}. The improvement brought by the phase-over-aperture model is substantial. But in order for it to become useful for image restoration, the speed needs to be further improved and the simulator needs to be differentiable.

To enable a differentiable simulator while being fast, people recognized that the blur can be efficiently represented in a low-dimensional space using fixed and pre-defined basis functions, as follows:
\begin{equation}
\vh_i = \sum_{\ell=1}^L \beta_{i,\ell} \vvarphi_\ell.
\end{equation}
Here, the functions $\vvarphi_\ell$ is the $\ell$th basis function of the blur, and $\beta_{i,\ell}$ is the $\ell$th coefficient for the $i$th blur kernel. A common choice of the basis functions is the Gaussian basis and its derivatives, although the authors of \cite{Mao_2021_a} showed that one can also perform principal component analysis (PCA) to construct the basis functions.

\begin{figure}[h]
\centering
\includegraphics[width=\linewidth]{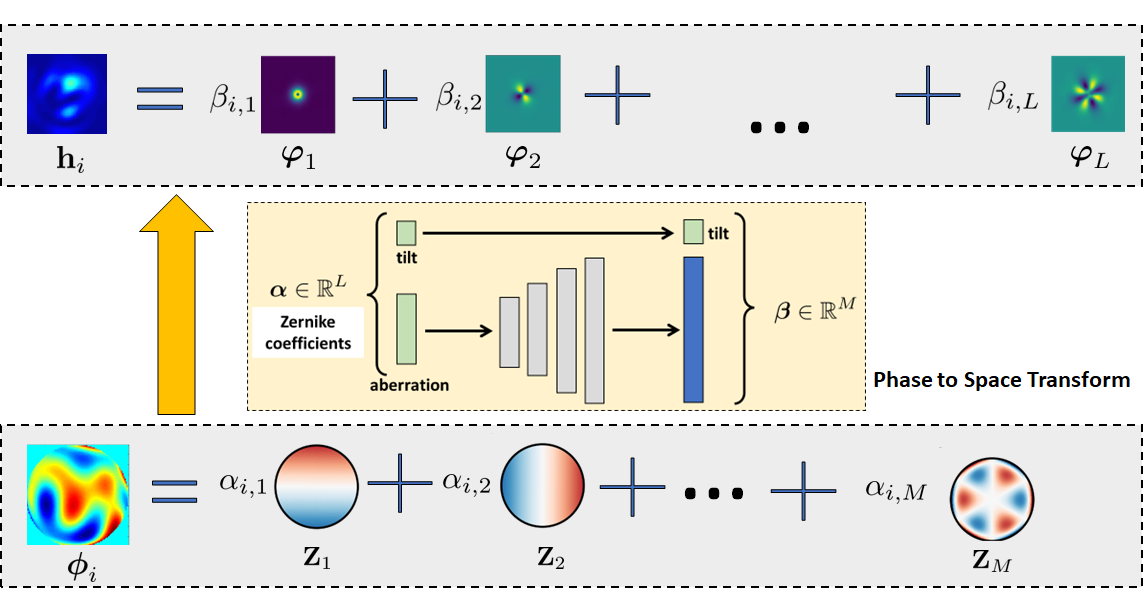}
\caption{Phase-to-space (P2S) simulator $\calH_{\vtheta}^{\text{P2S}}$ defines a mapping from the phase domain to the spatial domain using two different forms of linear representations. The mapping is implemented using a shallow neural network, and hence the model is differentiable.}
\label{fig: P2S}
\end{figure}

Why do we want to write $\vh_i$ as a linear combination of the basis functions? The reason is that we can now represent all the $\vh_i$'s using the coefficient vector $\vbeta = \{\beta_{i,\ell} \,|\, i = 1,\ldots,N, \ell = 1,\ldots,L\}$. Then, as long as the relationship between $\valpha$ and $\vbeta$ is established, we can go back and forth between the representations.
\begin{equation}
\valpha
\overset{\text{\tiny{Phase to Space}}}{
\longrightarrow}
\vbeta
\end{equation}
The forward mapping from $\valpha$ to $\vbeta$ is called the phase-to-space transform (P2S) \cite{Mao_2021_a} \footnote{The inverse mapping from $\vbeta$ to $\valpha$ is known as the space-to-phase transform (S2P). The implementation of S2P remains an open problem, as of today.}. P2S is implemented using a very shallow neural network because the dimensionality of the input is typically $\valpha \in \R^{36}$ whereas that of the output is around $\vbeta \in \R^{100}$. Thus, a shallow 3-layer fully connected network is sufficient to learn the mapping. \fref{fig: P2S} illustrates the conceptual diagram of the P2S.

With P2S, the turbulence model is now fully differentiable. This is because the distorted pixel can be obtained through the linear equation
\begin{equation}
\widehat{\vy}_i =
\underset{\calH_{\vtheta}^{\text{P2S}}(\vx_i)}{\underbrace{\sum_{\ell=1}^L \beta_{i,\ell} \Big[\vvarphi_\ell \circledast \vx\Big]}},
\end{equation}
where $\vtheta = \vbeta$. As long as $\vbeta$ is known, the simulator only needs to perform the basis convolutions. Since $\vbeta \leftrightarrow \valpha$ via P2S, the simulation model is completely characterized by the Zernike coefficient $\valpha$. Moreover, since the P2S model $\calH_{\vtheta}^{\text{P2S}}$ is linear in $\vbeta$, any gradient with respect to $\vbeta$ can be computed. Even if we need to compute the gradient with respect to $\valpha$, we can do so by backpropagating the gradient through the P2S neural network (which is a three-layer fully connected network.)

\boxedeg{
\vspace{1ex}
CIF Trade-off for \textbf{Phase-to-Space (P2S) Model}:
\begin{itemize}
\setlength\itemsep{0ex}
\item $\calE_{\text{sim}}$: Same simulation error as $\calH_{\vtheta}^{\text{phase}}$. P2S is just a (much) faster version.
\item $\calE_{\text{complexity}}$: Lower complexity than the phase-over-aperture model.
\item $\calE_{\text{recon}}$: Small. Since $\calH_{\vtheta}^{\text{P2S}}$ is significantly faster than $\calH_{\vtheta}^{\text{phase}}$, it can generate a much larger dataset. $\calH_{\vtheta}^{\text{P2S}}$ can be integrated into the reconstruction model.
\item $\calE_{\text{diff}}$: Zero. The model is differentiable.
\end{itemize}
}

\textbf{Further Improvements}. There are additional things people proposed to improve the speed and generality of the P2S simulators. For example, in \cite{Chimitt_2022_a}, it was shown that the off-diagonal blocks of the correlation matrix $\E[\valpha\valpha^T]$ can be truncated without hurting the performance. In \cite{Chimitt_2023_b}, it was shown that with a carefully designed reformulation, the simulator can be generalized to an arbitrary turbulence profile along the path (i.e., the structure constant $C_n^2$ changes with respect to the distance). Both changes do not affect the differentiability of the simulator, but they make the simulator even faster and more accurate. Another modification is the recognition of the scattering/gathering forms of convolution in \cite{Chimitt_2023_a}. It was shown that for the physics to be valid, the convolution needs to be implemented in the scattering form.

\subsection{Simulator-in-Reconstruction}
An important aspect of CIF is to use simulators in reconstruction. The goal of this subsection is to explain how this can be (has been) achieved.

\textbf{Simulator as Synthesizer for Training Data}. The most straightforward application of the simulators is to use them to synthesize training data. Starting with a collection of clean images, the simulator generates the turbulence effect to produce a distortion-clean pair.

One often overlooked usage of a simulator is its \emph{flexibility} in synthesizing data according to the needs of the training:
\begin{itemize}
\item Multiple scales: A simulator can synthesize turbulence across different resolutions. For example, for a fixed propagation distance and object size, a lower resolution image would require a smaller displacement vector and a smaller blur kernel. These images are often easier to restore. If the simulator can produce multi-resolution distortions, the overall restoration will be benefited.
\item Tilt-free, blur-free, all-free: A simulator can generate tilt-only distortions, blur-only distortions, or no distortion. This provides a powerful way to \emph{disentangle} the coupling effects of the turbulence and object movement. For approaches such as knowledge distillation or student-teacher learning, the decoupling capability offered by a simulator is the key enabler because there is no alternative ways we can train those models.
\end{itemize}

\begin{figure}[h]
\begin{tabular}{cc}
\multicolumn{2}{c}{\includegraphics[width=0.95\linewidth]{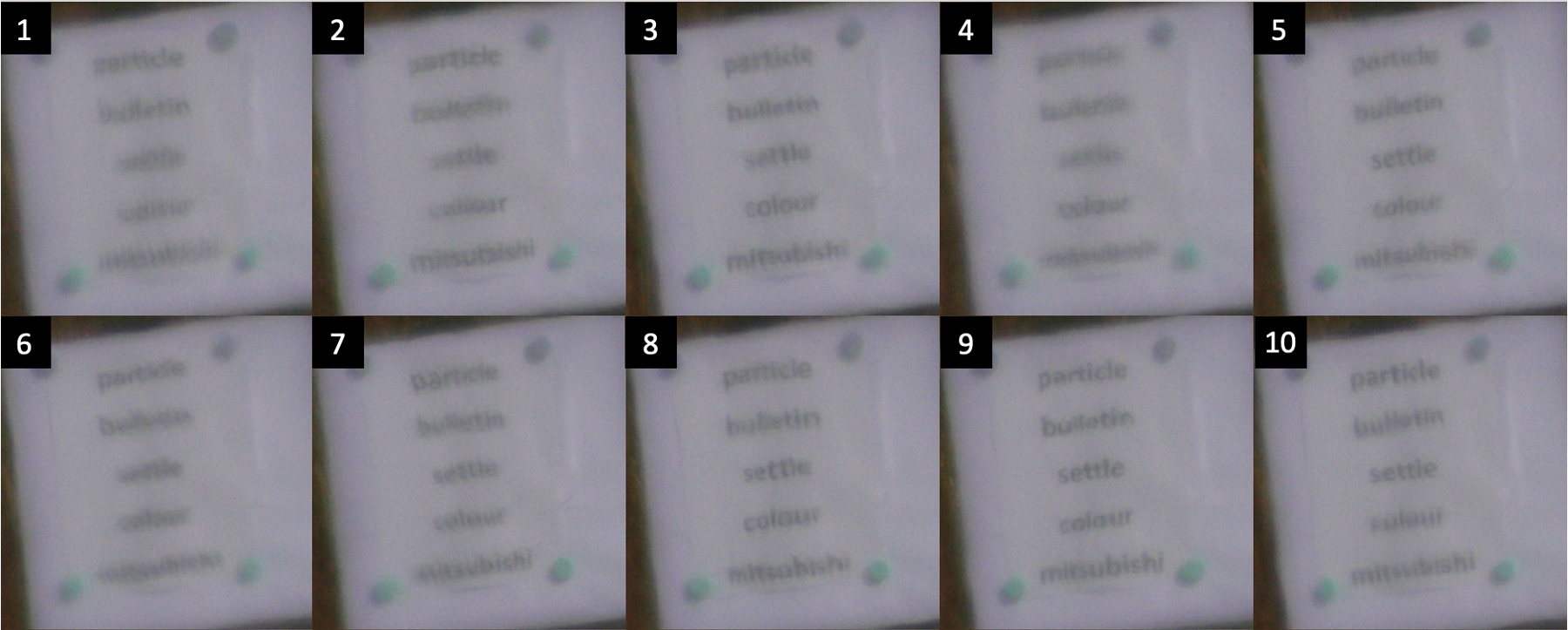}}\\
\multicolumn{2}{c}{\small{Input frames}} \\
\includegraphics[width=0.46\linewidth]{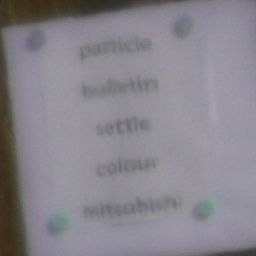}&
\hspace{-2ex}
\includegraphics[width=0.46\linewidth]{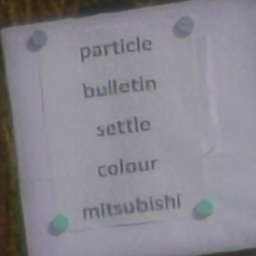}\\
\small{(a) $\calR_{\vpsi}$ trained with} & \small{(b) $\calR_{\vpsi}$ trained with} \\
\small{data synthesized by } & \small{data synthesized by }\\
\small{TSRWGAN \cite{Jin_2021_TSRWGAN}} & \small{P2S + variants \cite{Mao_2022_a, Zhang_2022_a}}
\end{tabular}
\caption{How much difference does a simulator make? This figure shows the reconstruction result of a network trained using datasets generated by two different simulators: TSRWGAN \cite{Jin_2021_TSRWGAN} and P2S \cite{Mao_2022_a}. With a fixed network architecture, the effect of the simulator is evident.}
\label{fig: simulator difference}
\end{figure}

How much improvement does a good simulator offer when compared to a bad simulator? Various papers have reported a consistent observation that the difference is significant \cite{Mao_2022_a, Zhang_2022_a}. \fref{fig: simulator difference} shows a comparison between TSRWGAN \cite{Jin_2021_TSRWGAN} and a variant of the P2S \cite{Mao_2022_a}, reported in \cite{Zhang_2022_a}. A common neural network architecture is chosen, and it is trained using two different datasets. TSRWGAN is a more rudimentary simulator with simple deformable grid and blurs, whereas P2S is more advanced. The results shown in \fref{fig: simulator difference} provides a strong evidence that a better simulator indeed makes a big difference in terms of image restoration.

\textbf{Simulator inside a Generative Adversarial Network}. A simulator can be directly used in image restoration, for example through the generative adversarial network (GAN) \cite{Lau_2021_b, Feng_2022_TurbuGAN}. In the GAN setting, the simulator can be used as part of the generative branch to synthesize what a distorted image should look like. This mirrors the nature where the image formation is determined by physics and the image sensor. The performance of the simulator has a direct impact to the performance of the GAN. If the simulator fails to mimic nature, then the simulated distorted image would not appear similar to the true distorted image. This puts more burden to the generator where it needs to compensate for the mismatch error caused by the simulator in addition to generating the latent unknown image. An accurate split-step propagation would not work either because it is simply too slow and it is not differentiable.

\textbf{Simulator for Re-degradation Loss}. Besides GAN, another approach to use simulator in the reconstruction loop is to consider the consistency loss, defined as
\begin{align}
\text{Consistency}\Big(\underset{\calH_{\vtheta}(\widehat{\vx})}{\underbrace{\calH_{\vtheta}(
\calR_{\vpsi}(\vy)}}
), \; \calG(\vx)\Big).
\end{align}
This is an additional loss put on top of the traditional reconstruction loss $\text{ReconLoss}(\widehat{\vx}, \vx)$.

\begin{figure}[h]
\centering
\includegraphics[width=\linewidth]{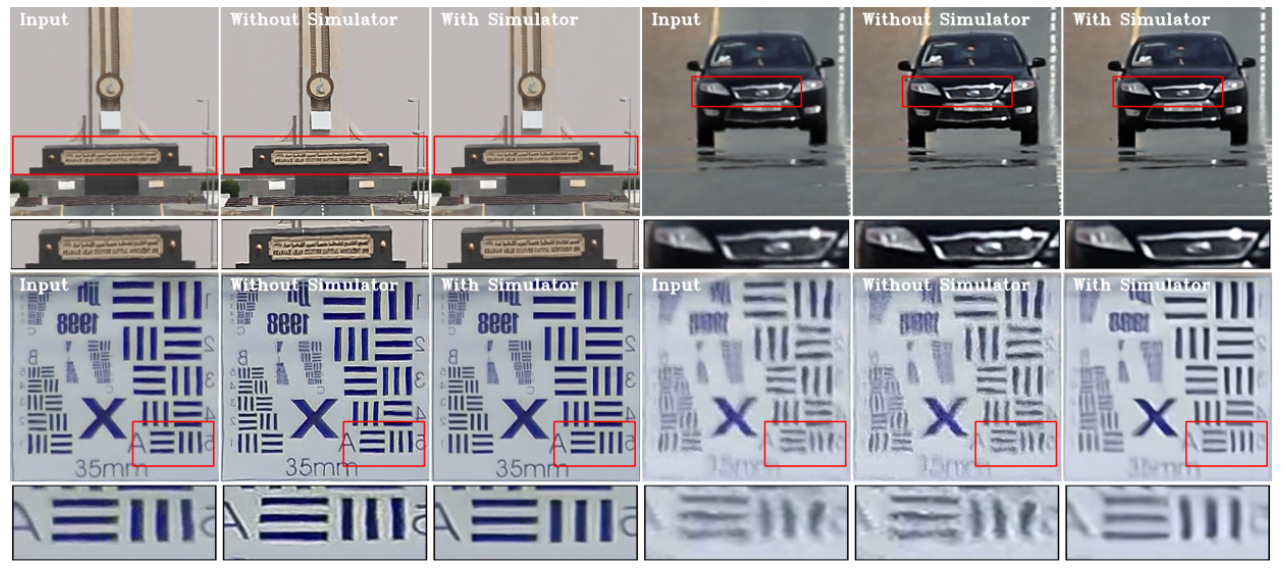}
\caption{What happens when the simulator is not used? The image reconstruction performance will be significantly worse. The figure is adopted from \cite{Jaiswal_2023_ICCV}.}
\label{fig: PiNet result}
\end{figure}

The performance of the image reconstruction depends on two factors: (i) How good is $\calH_{\vtheta}$ compared to $\calG$? (ii) How good is $\widehat{\vx}$ compared to the ground truth $\vx$? Many people ask why $\text{ReconLoss}(\widehat{\vx}, \vx)$ is insufficient. The answer is that the reconstruction loss never provides any explicit knowledge about the forward model. Since $\calG$ and $\calH_{\vtheta}$ are often ill-conditioned and so many $\vx$ can be mapped to the same $\vy$, the consistency loss helps the reconstruction by enforcing it not to create artifacts that cannot be explained by the forward model. The benefit of the consistency loss is supported by numerical evidence in \fref{fig: PiNet result}.

\textbf{Looking Forward.} The results of the latest reconstruction methods are promising. However, in my opinion, the full power of CIF is yet to be explored. Here are a few observations:
\begin{itemize}
\item \textbf{State vector $\vtheta$ update}. As I explained in Section E.2, the state vector update is analogous to the blur kernel estimation problem in blind deconvolution. Given a distorted image, having the ability to estimate the turbulence parameters could significantly reduce the uncertainty of the reconstruction. However, as of today, these ideas are yet to be developed.
\item \textbf{State encoding}. The state vector $\vtheta$ is often a high dimensional vector, e.g., the collection of Zernike coefficients over the entire image. However, it is likely that $\vtheta$ has a low dimensional representation. As of today, there is little understanding of how these state vectors can be encoded more efficiently.
\item \textbf{Bijective mapping from pixel space to embeddings}. An open problem today is the non-uniqueness of the low-dimensional representation. Using Zernike coefficients as an example, it is possible that two sets of Zernike coefficients can give the same pixel-level image distortion. Therefore, while phase-to-space is easy to do, the inverse mapping space-to-phase can be significantly harder. For many problems dealing with optics through different environments, such a bijective mapping is an important technical challenge.
\item \textbf{Ultra-fast simulator}. Advanced simulators such as P2S can achieve 40 frames per second for a 512-by-512 image. However, for P2S to be used as part of an iterative algorithm or used as an integral part of the reconstruction neural network, the runtime would probably need to be suppressed to a microsecond range. This is a major challenge for both hardware (GPU) and algorithm.
\end{itemize}

\section{More Examples}
In this section, I would like to elaborate more on how CIF could fit other imaging problems by discussing a few other examples in addition to the turbulence example above.

\subsection{De-raining and de-hazing}
Consider the problem of imaging through rain and haze. Like imaging through atmospheric turbulence, the exact ground truth image formation $\vy = \calG(\vx)$ cannot be determined exactly due to the stochastic nature of the process \cite{Cantor_1978, Garg_2005_ICCV}.

In the literature, the degradation can be modeled in different ways \cite{Li_2019_CVPR}. For example, if the distortion is caused by \textbf{rain streak}, then
\begin{equation}
\calH_{\vtheta}^{\text{streak}}(\vx) = \vx + \mathbf{b},
\end{equation}
where $\mathbf{b}$ is a sparse vector representing the line streak effect of the rain. If the distortion is caused by \textbf{adherent raindrops}, \cite{You_2016_PAMI} proposes a model
\begin{equation}
\calH_{\vtheta}^{\text{raindrop}}(\vx) = (1-\mM) \odot \vx + \vd,
\end{equation}
where $\mM$ is a binary mask indicating whether a pixel has a raindrop, $\odot$ is the elementwise multiplication, and $\vd$ is a sparse vector with localized scattering raindrops. If the distortion is caused by \textbf{haze and rain streak}, then the model is
\begin{equation}
\calH_{\vtheta}^{\text{haze}}(\vx) = \vx \odot \vt + \mA (1-\vt) + \mathbf{b},
\end{equation}
where $\vt$ is the transmission map (see Example 2), $\mA$ is the airlight color transformation, and $\mathbf{b}$ is a sparse vector representing the streak.

In recent years, there is an increasing amount of efforts to use neural networks to \emph{learn} the model so that $\calH_{\vtheta}$ can be more similar to $\calG$. For example, \cite{Yue_2021_CVPR} proposes the model
\begin{equation}
\vy_t = \calH_{\vtheta}^{\text{dynamic}}(\vx_t) = \vx + \mathbf{b}_t + \vn_t,
\end{equation}
where $t$ denotes the time, and $\vn_t \sim \text{Gaussian}(0,\sigma^2)$ is the noise vector. The rain model $\mathbf{b}_t$ is defined through some variations of the Markov process such that
\begin{align}
\vb_t &= \calA_{\valpha}(\vb_{t-1}),
\end{align}
where $\calA_{\valpha}$ is a neural network that provides a memoryless update based on previous time stamps. Because $\calA_{\valpha}$ is a neural network, it inherits several desired properties such as being differentiable. Since the simulator is co-optimized with the reconstruction algorithm, the reconstruction performance is evidently better.

Besides these examples, the general direction of imaging through adverse weather today is to inject more physics into the problem \cite{Kadambi_2023_Nature}. In the area of rain, snow, and fog, there is an increasing amount of high quality physics-based simulators that can simulate these optical effects \cite{Ba_2022_ECCV, Zhang_2023_CVPR}. The usage highlights the relevance of CIF.

\subsection{Differentiable optics}
\begin{figure}[h]
\centering
\includegraphics[width=\linewidth]{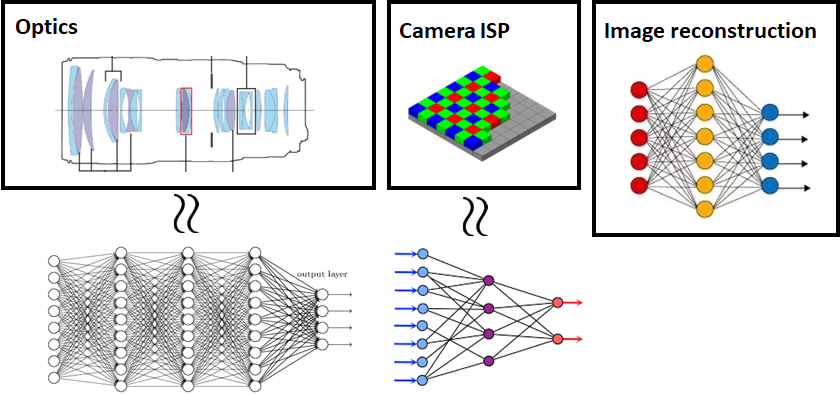}
\caption{Differentiable optics. In a conventional camera system, there is an optical component, an image signal processing (ISP) unit which controls the exposure, color filter, etc, and an image reconstruction algorithm. Differentiable optics aims at approximating the physical optics module and the ISP with neural networks so that the entire system is differentiable. This will enable end-to-end training of the image reconstruction algorithm.}
\label{fig: dO}
\end{figure}

While CIF is mostly concerned with degradation processes arising from nature, the concept can be applied to other forms of optics-algorithm co-design problems such as the one summarized in \fref{fig: dO}.

(1) \textbf{Differentiable optics for lens systems} \cite{Tseng_2021_TOG, Wang_2022_TCI, Chen_2023_APR, Sitzmann_2018_SIGGRAPH}. Traditional lens design is a standalone process where people use ray-tracing tools such as Zemax to optimize the parameters of the lenses. If one wants to design the downstream image reconstruction algorithm, these ray tracing tools, however, would be incompatible with the reconstruction. To overcome the difficulty, various methods have been proposed to approximate the true lens system $\calG$ with neural networks $\calH_{\vtheta}$. Since the reconstruction is usually a neural network, having a neural network $\calH_{\vtheta}$ will give us an end-to-end differentiable camera system.

Today, $\calH_{\vtheta}$ is mainly used to improve the image reconstruction. There is relatively little work on co-designing the lens parameters. The reason is that co-designing the lens parameters would require a method to ``translate'' the weights of the neural network $\calH_{\vtheta}$ to the lens parameters. This is largely an open problem.

(2) \textbf{Metasurface design} \cite{Dean_2022_arXiv,Chen_2018_Nature}. Another usage of CIF is the design of the metasurfaces. Metasurfaces are nanoscale materials where each element can be engineered to perform a specific phase operation. Compared to traditional glass-based lenses which are bulky, metasurfaces are substantially thinner while achieving a competitive optical performance. The design of the metasurfaces is often performed together with the image reconstruction algorithm. This is because metasurfaces today still have many limitations in terms of chromatic aberration control and spatially varying points spread functions. Therefore, it is necessary to co-design a deconvolution algorithm.

(3) \textbf{Differentiable rendering} or computational light transport \cite{Velten_2016_ACM, OToole_2010_TOG, Gkioulekas_2015_TOG}. The problem here is more concerned with a realistic rendering of objects in computer graphics. For example, as light propagates through milk and wax, how does the image look like? Or, if the light source is located at a certain position in the room, how will the light bounce among the walls and eventually reach the camera? Because of the complexity of the actual environment and the underlying image formation process (which sometimes requires us solving partial differential equations), newer approaches attempt the problem by approximating the ground truth $\calG$ with a neural network $\calH_{\vtheta}$.

\subsection{Image sensor circuit model}
Thus far I have been mainly talking about optics. But CIF can be extended to other components such as the circuit level modeling of image sensors from photo diodes, comparator, capacitors to the output signal.

\begin{figure}[h]
\centering
\includegraphics[width=\linewidth]{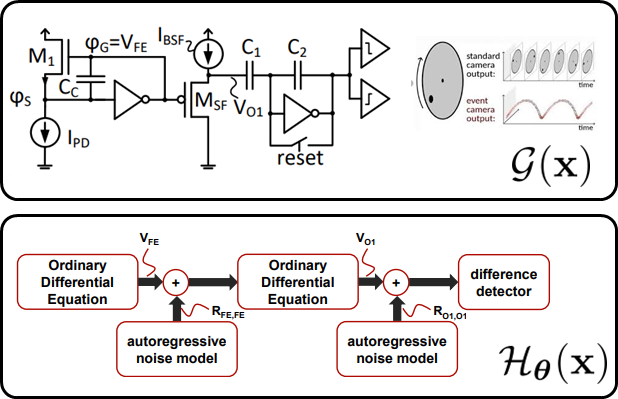}
\caption{What happens when the simulator is not used? The image reconstruction performance will be significantly worse. The figure is adopted from \cite{Suess_2023_IISW}.}
\label{fig: DVS}
\end{figure}

The top row of \fref{fig: DVS} shows the circuit diagram of a dynamic vision sensor (DVS) \cite{Suess_2023_IISW}. The exact signal formation process $\vy = \calG(\vx)$ is both stochastic and complicated. However, it is possible to approximate the transient behavior using an ordinary differential equation and the probabilistic events by drawing samples from a pre-defined covariance matrix. This leads to the bottom row of \fref{fig: DVS} where there are two ordinary differential equation blocks and two autoregressive blocks. Simpler models have been proposed \cite{Mou_2022_EI}, but it was shown that the performance is not as good.

Besides DVS, there is also a growing interest in novel digital image sensors including photon counting devices, quanta image sensors (QIS), and single-photon avalanche diodes. Using QIS as an example, a series of 1-bit and multi-bit models have been proposed \cite{Chan_2022_OneBit, Elgendy_Threshold_2018, Gnanasambandam_TCI_HDR} and analyzed \cite{Chan_SNR_2022, Chan_Density_2022}, together with various image reconstruction algorithms \cite{Chan_MDPI_2016, Chi_ECCV_2020, Gnanasambandam_Megapixel_2019, Choi_ICASSP_2018, Elgendy_Demosaic_2021} and applications \cite{Gnanasambandam_ECCV_2020, Chengxi_ICCVW_2021}.

\section{Thoughts and Discussions}
In this section I summarize a few commonly asked questions about CIF.

\textbf{Is CIF = model-based image reconstruction (MBIR)?} My view is that CIF and MBIR are aiming for two different goals. In MBIR, the premise of the problem is that someone has given us a model $\widehat{\vy} = \calH_{\vtheta}(\vx)$. Our job is to find the best algorithm to solve for $\vx$, by exploiting various signal priors such as sparsity or generative models. CIF, in contrast, primarily focuses on the design of $\calH_{\vtheta}$. As I illustrated in the previous sections, there are accurate but complex $\calH_{\vtheta}$ and less accurate but effective $\calH_{\vtheta}$ for the inverse solver. Constructing a meaningful $\calH_{\vtheta}$ while maintaining the computational efficiency is what CIF is about.

\textbf{Why not just use neural networks to approximate $\calG$?} With the growth of building deep neural networks to mimic the optics, we might be tempted to think that a good CIF simulator must be a neural network (so that everything is differentiable). I personally think that this is not the best (nor the only) direction. While I completely acknowledge the power of a neural network, I do not think today's neural networks have advanced to the stage that it can perform every task without using a much larger model. I agree that for some specialized tasks such as solving an ordinary differential equation, a neural network could offer a powerful approximation. But for  equations such as Fresnel propagation, Fourier transforms are much more efficient.

I envision that future CIF simulators will most likely be a hybrid model where neural networks are used as one of the building blocks to complete some specific sub-tasks.  Differentiability can be ensured \emph{without} a neural network. For example, in the turbulence case study I described above, the differentiability is enabled by a different representation of the phase function. Even for tasks such as ray tracing, it will be far more interesting (and impactful) to derive new equations that preserves differentiability without using automatic differentiation of computational graph of any sort.

\textbf{Is sensor-algorithm co-optimization always needed?} As I follow the past few year's of publications in computational imaging, I observe a trend that whenever we see a sensor and an algorithm, it will be sensor-algorithm co-optimization. I can see the necessity of co-optimization if the goal is to maximize the system's performance unconditionally. However, from a practical point of view, we should not forget about the feasibility and physical constraints. In a recent paper \cite{Qu_2023_SVE}, it was shown that co-optimization brings negligible benefits to the actual performance in some specific problems. This counter example is perhaps a good reminder to us about the reality.

\section{Conclusion}
Computational imaging is the intersection of image acquisition, signal prior, and numerical algorithm. Forty years ago when we were still in the beginning of solving inverse imaging problems, our attention was mostly spent on developing better and more powerful priors (i.e., regularization functions such as $\ell_1$ norm, total variation, Markov random field, etc) together with faster numerical algorithms (e.g., gradient descent, alternating minimization, operator splitting, etc). As we continue to advance computational imaging in 2023, it is perhaps time to rethink about the role of the \emph{forward model} that describes the image acquisition process.

This article describes the concept of computational image formation (CIF). CIF highlights the choice of a simulator $\calH_{\vtheta}$ that approximates the true image formation process $\calG$. Unlike a physics simulator whose goal is to match $\calG$ unconditionally, the simulator $\calH_{\vtheta}$ in CIF needs to maximize the image reconstruction performance while matching $\calG$ up to some level. Moreover, the simulator needs to be very fast so that it can be used to generate data, and it needs to be differentiable so that it can be used in the reconstruction loop. Several examples are elaborated to explain CIF.

We all stand on the shoulder of giants. CIF is no exception. It is a concept summarizing decades of collective efforts of the computational imaging community. As we look forward to the future of imaging, I envision that simulators will play a unprecedented role in the deep learning era.

\section*{Acknowledgement}
The research is based upon work supported in part by the Intelligence Advanced Research Projects Activity (IARPA) under Contract No. 2022‐21102100004, and in part by the National Science Foundation under the grants 2133032, 2030570, 2134209 and 1763896. The views and conclusions contained herein are those of the authors and should not be interpreted as necessarily representing the official policies, either expressed or implied, of IARPA, or the U.S. Government. The U.S. Government is authorized to reproduce and distribute reprints for governmental purposes notwithstanding any copyright annotation therein.

\balance
\bibliographystyle{IEEEtran}
\bibliography{egbib}


\begin{biography}
Stanley Chan received his PhD in electrical engineering from UC San Diego (2011). He is currently an Elmore Associate Professor of ECE at Purdue University, West Lafayette. His research interests includes computational imaging, single-photon image sensors, and imaging through turbulence. He is a senior area editor of IEEE Transactions on Computational Imaging, and a recipient of the 2022 IEEE Signal Processing Society Best Paper Award.
\end{biography}

\end{document}